\documentclass[]{interact}

\usepackage{epstopdf}% To incorporate .eps illustrations using PDFLaTeX, etc.
\usepackage[caption=false]{subfig}% Support for small, `sub' figures and tables
\usepackage{setspace}
\usepackage{mhchem}
\usepackage{xcolor}
\usepackage[numbers,sort&compress]{natbib}% Citation support using natbib.sty
\bibpunct[, ]{[}{]}{,}{n}{,}{,}% Citation support using natbib.sty
\renewcommand\bibfont{\fontsize{10}{12}\selectfont}% Bibliography support using natbib.sty

\theoremstyle{plain}% Theorem-like structures provided by amsthm.sty

\theoremstyle{definition}

\theoremstyle{remark}

\begin{document}

\articletype{ARTICLE TEMPLATE}% Specify the article type or omit as appropriate

\title{Three-dimensional Vorticity Effects on Extinction Behavior of Laminar Flamelets}
%\maketitle
\author{
\name{Wes Hellwig\thanks{CONTACT Wes Hellwig. Email: whellwig@uci.edu} and Xian Shi and William~A. Sirignano}
\affil{Department of Mechanical and Aerospace Engineering, University of California Irvine, Irvine, California, USA}
}

\maketitle

\begin{abstract}
A recent rotational flamelet model (Sirignano \cite{Sirignano1,Sirignano2,Sirignano3}) is developed and tested with an improved framework of detailed chemistry and transport. The rotational flamelet model incorporates the effects of shear strain and vorticity on local flame behavior and is three-dimensional by nature. A similarity solution reduces the three-dimensional governing equations to ODEs involving a transformation to a non-Newtonian reference frame. A 9-species chemical kinetics model is used for $\mathrm{H_2}$-$\mathrm{O_2}$ combustion with non-reacting $\mathrm{N_2}$. In all non-premixed flame cases, the oxidizer is pure $\mathrm{O_2}$ while the fuel $(\mathrm{H_2})$ is diluted with $\mathrm{N_2}$. Multiple flamelet cases including non-premixed, premixed, and partially-premixed flames are performed. Across all cases, vorticity extends flammability limits by up to 30\% in terms of the ambient extinction strain rate and modifies both local flame structure and mixture composition. For $\mathrm{N_2}$-diluted $\mathrm{H_2}$-$\mathrm{O_2}$ non-premixed flames, where the location of minimum density coincides with the location of peak temperature, the centrifugal force induced by vorticity reduces the mass flow rate through the flame, effectively lowering the local strain rate. This increases residence time, thus extending flammability limits and reducing burning rates. This analysis is done also for premixed and partially-premixed flames. For pure $\mathrm{H_2}$-$\mathrm{O_2}$ non-premixed flames, where minimum density lies between the flame zone and the fuel inlet boundary, centrifugal forces do not significantly modify flame behavior. Stable and unstable branches of S-curves for non-premixed and partially-premixed flames and stable branches for premixed flames show extended flammability limits due to vorticity. The capabilities of the rotational flamelet model reveal that vital physics are currently missing from two-dimensional, irrotational, constant-density, flamelet models. Improvements of detailed chemical kinetics, transport formulation, and thermo-physical properties bring the new flamelet model to par in these areas with existing models, while adding new features in terms of physical emulation.
\end{abstract}


\section{Introduction}
Combustion processes in gas turbine and rocket engines involve high mass fluxes and are inherently turbulent, making simulations of such processes challenging. For flows of these scales, the computation time of direct numerical simulations (DNS) is prohibitively costly, precipitating the need for large-eddy simulations (LES) or Reynolds-averaged Navier-Stokes (RANS) simulations. These techniques, as opposed to DNS, do not resolve all length scales of the turbulent flow, instead filtering smaller length scales below the mesh size (LES) or neglecting fluctuating components of flow variables (RANS), and modeling the unresolved terms. In the case of turbulent reacting flows, both the Reynolds stress and chemical source terms are unresolved. This paper is concerned only with the latter. 

Multiple methods exist to model the sub-scale chemical processes and can be generally categorized as flamelet or non-flamelet approaches. Here, we specifically aim to model laminar flamelets that are relevant to practical turbulent combustion. Our model is not aimed at laminar-flow laboratory experiments with a controlled counterflow issuing from nozzles. Examples of non-flamelet approaches include probability density function methods \cite{AnandPope,LindstedtVaos} and the conditional moment closure method \cite{KlimenkoBilger}.

The original flamelet theory, described by Williams \cite{Williams1975}, approximates turbulent flames existing in the diffusion layer between oxidizer and fuel interfaces as an ensemble of thin, highly sheared, one-dimensional, diffusive-reactive zones, where each may be approximated by a counterflow non-premixed flame or ``flamelet''. The ensemble of flamelets implies that any one flamelet exists only in a small section of a turbulent interface, thus avoiding the flawed assumption of chemical equilibrium throughout the flow. It is assumed that the Damköhler number, defined on the Kolmogorov scale, is sufficiently high, such that chemistry can instantaneously respond to changes in the flow. Thereby, flamelets are in quasi-steady states. Peters \cite{Peters} states that the effects of turbulent straining can be neglected if the flame scale is significantly smaller than the Kolmogorov scale. This condition is definitively upheld only for premixed flames in the wrinkled flamelet regime where no eddies exist within the flamelet. The premixed flame speed does not depend strongly on the applied strain rate. At lower values of applied strain rate, the premixed flame will be shown to exist outside of the central core of the eddy. In that case, the eddies can cause wrinkling of the flame front, although we will not elaborate on that aspect here. Later in the paper, we will show that, at sufficiently high strain rates, the premixed flame will move into the core of the eddy within the Kolmogorov scale. For non-premixed flames and partially-premixed flames, the diffusion layer and viscous layer are commensurate in size and determined by the applied strain rate and therefore the diffusion layer of the flame cannot be smaller than the Kolmogorov scale.

Veynante and Candel \cite{PoinsotVeynanteCandel} introduced a method for turbulent premixed combustion whereby knowledge of the turbulence integral scale and turbulent kinetic energy can determine whether the flow will feature classical flamelets, pockets, or distributed reaction zones, suggesting extended applicability of flamelet modeling beyond the classical theory. Premixed experimental studies \cite{Dunn1,Dunn2} support this, presenting evidence that thin, flamelet-like structures and thickened flames with increased turbulence interaction (which experience extinction and re-ignition), can both exist in highly turbulent flows. Attempts to capture these thickened premixed flames with increased combustion-turbulence integration exist \cite{Colin}. 

In the present work, we deal only with flamelets on the Kolmogorov scale, having the shortest characteristic length and time, such that we can make the quasi-steady assumption. The reader may question whether Kolmogorov eddies, having similar characteristic times to the flamelet diffusion time, would cause unsteadiness. Here, the quasi-steady assumption weakens, suggesting the need for unsteady flamelet analysis; however, important physical insights can still be gleaned with the quasi-steady assumption. There has been some development of an unsteady rotational flamelet model \cite{Sirignano2023wssci,Sirignano_unsteady}.

\subsection{Existing Models and Issues}\label{class}
The use of flamelets in LES and RANS provides a significant improvement in computational efficiency; however, the many assumptions on which existing flamelet theories are built result in limitations. Existing models (e.g., \cite{Pierce_Moin}) assume an axisymmetric or planar geometry and a corresponding strain field while frequent three-dimensionality has been observed \cite{Nomura1,Nomura2,Boratav1996,Boratav1998,Ashurst}. Three principal normal strain rates exist for such flows. For incompressible flow, one of the principal normal strain rates will be compressive, another will be tensile, and the third may be either compressive or tensile having an intermediate magnitude. In the counterflow configuration, compressive principal strain rates correspond to inflow while tensile principal strain rates correspond to outflow. Direct numerical simulations of non-premixed combustion in sheared turbulence indicate that vorticity most probably aligns with the intermediate tensile strain axis. For isotropic turbulence, vorticity is also likely to align with the intermediate principal strain axis but is less probable than in the sheared turbulence case \cite{Ashurst,Nomura1}. In a subsequent work \cite{Nomura2}, Nomura and Elghobashi performed DNS studies on non-premixed reacting flows and showed that vorticity can align with the maximum tensile axis but more often tends to align, for instances with higher strain rate, with the intermediate tensile axis. In a counterflow configuration, the diffusion plane is normal to the principal compressive axis; i.e., scalar gradients are aligned with the principal compressive axis \cite{Ashurst,Nomura1,Nomura2,Boratav1996,Boratav1998}. As the length scales decrease in turbulent flows, the magnitudes of shear strain and vorticity increase, suggesting that centrifugal forces could affect flamelets appreciably. Betchov \cite{Betchov} showed that a tensile intermediate principal strain rate is most critical to the generation of vorticity and the cascade of energy to smaller scales. In the same work, Betchov showed that the normal compressive strain rate is of the same order of magnitude as vorticity across all length scales {\cite{Betchov}}. This equivalency serves as the basis for choosing vorticity values, $\omega_k^*$, on the same order of magnitude as the asymptotic compressive strain rate, $S^*$, for our computations later in this manuscript. Some interesting more recent studies on DNS for premixed turbulent combustion are provided by Driscoll et al. \cite{Driscoll}, Chen and Im \cite{ChenIm}, Savard et al. \cite{Savard1} and Savard and Blanquart \cite{Savard2}. They address issues related to strain rate. These papers do not discuss relative alignments of vorticity, strain rate eigenvectors, and scalar gradients. Few recent DNS studies for non-premixed turbulent combustion exist. Since the studies of Ashurst et al. \cite{Ashurst}, Nomura and Elghobashi \cite{Nomura1,Nomura2}, and Boratav and Elghobashi \cite{Boratav1996,Boratav1998}, little attention has been given to vorticity in DNS studies for turbulent combustion.

Rotational effects are largely neglected in existing models. Karagozian and Marble \cite{Karagozian} treated a three-dimensional non-premixed flame with inward radial flow and outward axial jetting, including a vortex aligned with the jet axis. The presence of the vortex caused the flame sheet to wrap around the vorticity axis. This analysis makes an \textit{ad hoc} density correction however, not solving the momentum equation with variable density. Consideration of variable density compounds centrifugal effects and must not be neglected as certain models have done, e.g. \cite{Linan,Peters,Karagozian}. Colin et al. \cite{Colin} developed a turbulent premixed flamelet model for thickened flames, based on DNS studies of a pair of counter-rotating vorticies interacting with the flame front. This approach involves artificially thickened flames created by decreasing the pre-exponential factor of the Arrhenius law coupled with an efficiency function, derived from resolved scale quantities, to impart turbulent effects on the thickened flame. Additionally, classical flamelet theories assume, a priori, a diffusion or premixed flamelet while LES results indicate premixed, partially premixed, diffusion, and multi-branched flames can all appear at some time and some locations in a turbulent reacting flow \cite{Hamins,Sirignano2021a,LopezCamara,NguyenPopovSirignano,NguyenSirignano2018,NguyenSirignano2019}. This has been addressed with the introduction of multi-regime flamelet models \cite{KnudsenPitsch1,KnudsenPitsch2} which account both for premixed and non-premixed combustion. These models determine burning regimes using the flame index originally proposed by Yamashita et al. \cite{Yamashita} and involve a transformation with a variable independent of mixture fraction, allowing the use of the progress variable flamelet model (FPV).

The most utilized of the existing flamelet models are those derived from, and including, the progress variable model (FPV), pioneered by Pierce and Moin \cite{Pierce,Pierce_Moin}. The progress variable model provides a simple and unique method for coupling flamelet outputs to the resolved scale. Flamelets are generated and parameterized via mixture fraction and stoichiometric scalar dissipation rate (SDR) which are then mapped to the resolved scale via a presumed probability density function, which itself, is parameterized by mixture fraction, SDR, and the progress variable. The progress variable may be thought of as a quantity indicating the extent of reaction, typically taken to be a weighted sum of specific product mass fractions. Thus, a high progress variable indicates significant combustion and vice versa. Governing equations for the progress variable, mean mixture fraction, and mixture fraction variance are added to the resolved scale. When all components of the theory are combined, the progress variable is essentially indicative of the stoichiometric SDR through the mapping procedure. Thus, by knowing the resolved scale value of the progress variable and the mean and variance of the mixture fraction, a unique flamelet solution is determined.

In the FPV, Pierce and Moin \cite{Pierce_Moin} included the upper and middle branches of the maximum temperature versus SDR curve, which addressed the issue of unsteady combustion behavior such as extinction and re-ignition. Pierce and Moin also introduced the progress variable (which tracks the global progression of the locally reacting mixture as a function of scalar dissipation rate at maximum flamelet temperature), greatly simplifying computations by governing this variable with a partial differential equation on the resolved scale. The inputs to the FPV model however, namely the progress variable and the mixture fraction, are taken directly from the resolved flow to the flame zone in the flamelet. The present authors believe that, when treating turbulent combustion at the Kolmogorov scale, the resolved flow inputs should be scaled to free stream or ambient inputs at the flamelet scale, allowing the sub-scale physics to determine the velocity and scalar fields in the flame zone. In other words, the extinction of the flamelet is still the result of the local strain rate in the flame zone, but this local strain rate is determined from the local physics of the problem rather than scaled directly from the resolved scale and presumed via a PDF. While a significant improvement to flamelet modeling, the FPV has drawn certain criticisms with the arbitrary definition of progress variable. Furthermore, another approximation \cite{Pierce} is sometimes made whereby the flame is determined by diffusion alone (advection is neglected). By neglecting shear strain, it further assumes irrotationality. 

Flamelet approaches have considered variable density, using a modified determination of stoichiometric SDR \cite{KimWilliams1993}, while others have used counterflow non-premixed flame solutions generated with CHEMKIN \cite{LAPOINTE202094,Breda,Smallwood,Ma} and Cantera \cite{Pantano,Rieth,Proch,Groth} either as the flamelet solution or to validate an improved numerical method. However, both CHEMKIN and Cantera one-dimensional counterflow non-premixed flame models suffer from the imposition of nozzle inlets, making the ambient strain rate at the flamelet scale zero. That ambient strain rate is clearly non-zero because velocity derivatives increase in magnitude through the eddy cascade. The flamelet generators in Cantera and CHEMKIN also make no attempt to apply vorticity or transverse strain to the counterflow, limiting solutions to two-dimensional planar and axisymmetric geometries \cite{Cantera,CHEMKIN}.

\subsection{Introduction of a New Model}
Via appropriate scaling for turbulent flows, estimates for the Kolmogorov scales for velocity, length, time, strain rates and vorticity can be obtained. Strained non-premixed flames (i.e., counterflow flames) will not occur at smaller scales because higher strain rates than the Kolmogorov scale are not possible. Flames with thickness much smaller than an eddy size would be wrinkled flames propagating through the eddies rather than flamelets held in capture by an eddy. Premixed flames with size larger than some eddies would rely on turbulent diffusion to assist propagation. These flames might be large enough to describe well at the resolved scale. Our focus is on flamelets captured by the small eddies. Therefore, the physics of the flamelet can be approached through laminar modeling. We expect the diffusive-advective balance for these flamelets at approximately the Kolmogorov scales.

Sirignano \cite{Sirignano1,Sirignano2,Sirignano3} addressed these issues with a new flamelet model, outlining five foundational criteria that the improved model must abide by. These are: 1) the model should have the ability to compute non-premixed, premixed, and partially-premixed flames; 2) the effects of shear strain and vorticity should be considered; 3) the shear strain and vorticity applied in the sub-grid model should be determined from the resolved scale without a contrived progress variable; 4) the model should be three-dimensional and not limited to axisymmetric or planar cases; and 5) variable density should be considered \cite{Sirignano1}. Sirignano \cite{Sirignano1,Sirignano2} demonstrated this new model using one-step chemical kinetics for propane-oxygen combustion, Fickian diffusion, and a similarity solution in the non-Newtonian reference frame that reduces the three-dimensional governing equations to a set of ODEs. The inclusion of centrifugal force caused by vorticity in the momentum equations was shown to reduce mass flux into the counterflow by creating a more adverse pressure gradient. Reduction of the in-flowing mass fluxes increases the residence time which reduces the burning rate but extends flammability limits. Additionally, the centrifugal force alters the out-flowing directions of gas based on density. The lighter gas (the product of combustion) tends to flow out along the vorticity axis because its low inertia is overpowered by the centrifugal force. Conversely, higher-density gas (unburned propellants and some low-temperature combustion products) flows outward in the plane affected by vorticity. 

The long-term goal is to develop a flamelet model that will provide chemical closure for RANS and LES simulations of turbulent reacting flows. We address flames with the diffusive-advective balance at those Kolmogorov scales. In addition to scaling velocity derivatives from the resolved scale to the smaller length scales, rules are needed to scale the gradients of the scalar properties. These gradients will statistically increase as the turbulence wavenumber increases. That portion of the coupling is one-way. The flamelet model would produce a heat release rate based on these inputs. That heat release affects all scales and creates a two-way coupling between the resolved flow and the small eddy scale. A primitive attempt at coupling \cite{Sirignano1} was made using a mixing-length turbulence model for the resolved scale. A major challenge for advancement is to develop improved methods for scaling the gradients of scalar properties such as temperature and mass fractions.

The goals of the current study are to improve the three-dimensional rotational flamelet model by: 1) using multi-step chemical kinetic models to calculate chemical source terms; 2) using the multi-component diffusion formulation instead of binary Fickian diffusion; 3) treating the governing equations in their full form, i.e. making no assumptions of constant or proportional thermo-physical properties; 4) developing both the stable and unstable branch of the flammability curves; 5) providing a more detailed description of the dependence of flammability limits on normal strain rates and vorticity; and 6) comparing nozzle counterflow geometry to free stream counterflow geometry. With these improvements, we show results for non-premixed, premixed, and partially-premixed flames. In the present study, $\mathrm{O_2}$ is used as the oxidizer while $\mathrm{H_2}$ (diluted with $\mathrm{N_2}$ for non-premixed flames) is used as the fuel. Nitrogen is not considered as a reacting species.

The analytical and computational analysis as well as the comparison with nozzle-based counterflows is presented in Section 2. The results for non-premixed flames, premixed flames, and partially-premixed flames are given in Section 3, and concluding remarks are given in Section 4.

\section{Materials and Methods}
\subsection{Formulation and Governing Equations}
\begin{figure}[!htb]
\centering
\noindent\makebox[\textwidth]{%
\includegraphics[width=1.0\textwidth]{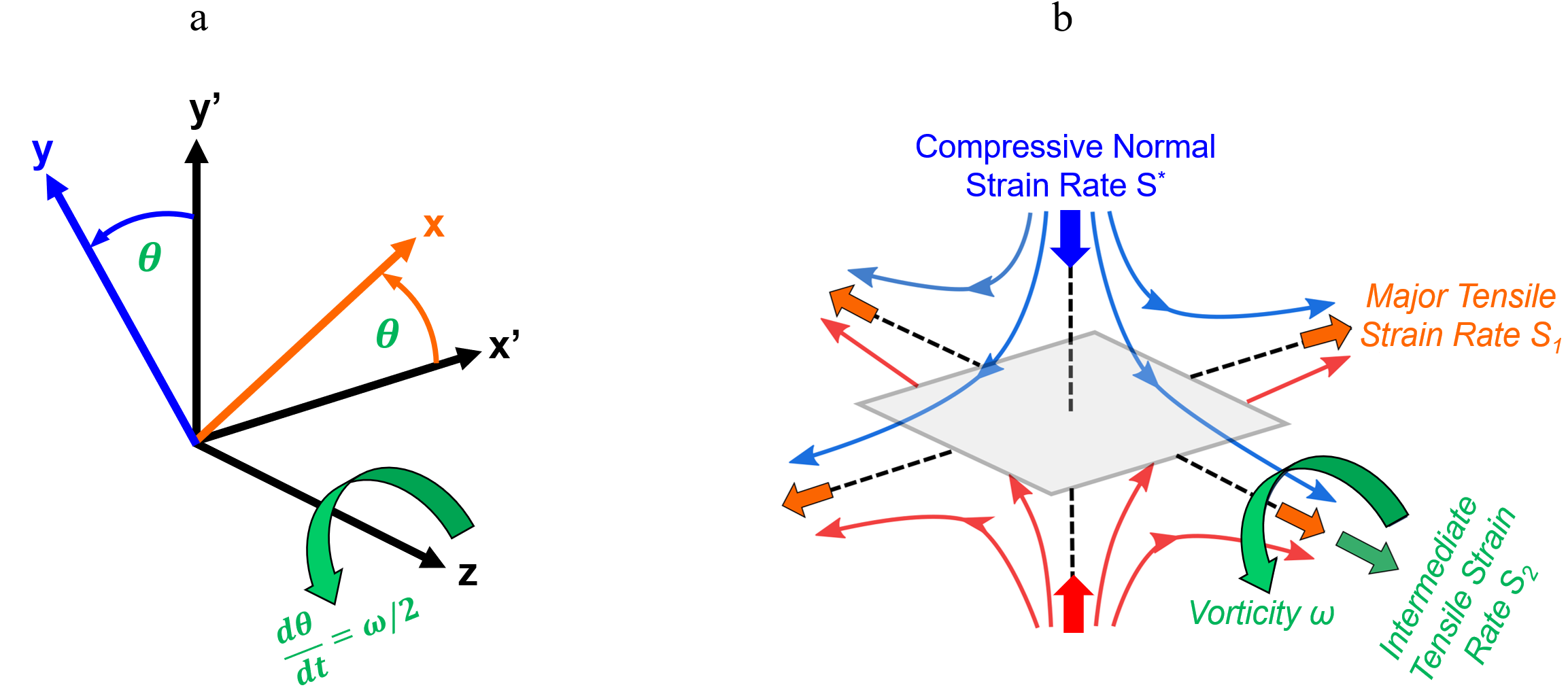}}
\caption{Schematic of the transformation to the non-Newtonian reference frame and alignment of strain rates and vorticity. (\textbf{a}) The black coordinate system shows the Newtonian reference frame for Eqs. (1-5). A non-Newtonian transformation is made such that \textbf{x'}$\rightarrow$\textbf{x} and \textbf{y'}$\rightarrow$\textbf{y} while \textbf{z} remains the same; i.e., the \textbf{x}-\textbf{y} plane rotates about \textbf{z} at the angular velocity $d\theta/dt = \omega/2$ imposed by the vorticity ($\omega$); (\textbf{b}) The counterflow is shown in the non-Newtonian reference frame (used for Eqs. (6-9)) with the compressive and tensile strain rates and vorticity aligned with their respective axes; non-premixed oxidizer side (blue), non-premixed fuel side (red). The reader should interpret this image as a snapshot in time as the counterflow structure, $S^*$, and $S_1$ will rotate about the \textbf{z} ($\omega$ and $S_2$) direction as time progresses.}
\label{fig1}
\end{figure}

Most flamelet approaches use scalar dissipation rate, a pseudonym for strain rate, at the non-premixed flame reaction zone (point of stoichiometric mixture fraction) as the key input parameter from which to determine scalar profiles. An a priori approximation relating strain rate to scalar dissipation rate across the flamelet is also frequently made. For example, Peters \cite{Peters} obtains a convenient Gaussian shape for scalar dissipation by inconsistently taking constant density in the momentum equation while using Chapman’s approximation for variable density in the scalar equation. In our description, turbulence imposes the strain rate and vorticity on the incoming flow while heat release within the eddy causes density change and the related modification of velocity and strain rate for the flow through the flamelet. So, the vorticity and strain rate of the far approaching fluid are the inputs to the analysis. Then, through momentum and mass conservation principles, the velocity and strain rates through the flamelet can be determined. The concept clearly parallels the standard for imposing conditions on the scalar properties; namely, values are imposed on temperature and composition for the upstream incoming flow and not in the flame zone. Figure \ref{fig1} \textbf{b} shows the three-dimensional geometry and associated strain rates. 

The governing equations for unsteady 3D flow, shown below, retain the non-Newtonian reference frame used by Sirignano; the vorticity vector aligns with the $z$-axis while the $\xi$ and $\chi$ axes rotate about $z$ at a constant angular speed, $d\theta/dt = \omega_k/2$ where $\omega_k$ is the Kolmogorov vorticity. See Figure \ref{fig1} \textbf{a} for the coordinate transformation.
\begin{equation}
    P = \rho R_{sp} T
\end{equation}
\begin{equation}
    \frac{\partial \rho}{\partial t} + \frac{\partial (\rho u_j)}{\partial x_j} = 0
\end{equation}
\begin{equation}
    \rho\frac{\partial u_i}{\partial t} + \rho u_j\frac{\partial u_i}{\partial x_j} + \frac{\partial p}{\partial x_i} = \frac{\partial \tau_{ij}}{\partial x_j} + \rho a_i
\end{equation}
\begin{equation}
    \frac{\partial (\rho h)}{\partial t} + \frac{\partial}{\partial x_j}\left(\rho u_j h\right) + \frac{\partial}{\partial x_j}\sum_{m=1}^N \rho v_{m,j}Y_m h_m - \frac{\partial p}{\partial t} - u_j\frac{\partial p}{\partial x_j} = \frac{\partial}{\partial x_j}\left(\lambda\frac{\partial T}{\partial x_j}\right) - \rho\sum_{m=1}^N h_{f,m}\dot{\omega}_m + \tau_{ij}\frac{\partial u_i}{\partial x_j}
\end{equation}
\begin{equation}
    \frac{\partial (\rho Y_m)}{\partial t} + \frac{\partial}{\partial x_j}\left(\rho Y_m u_j\right) + \frac{\partial}{\partial x_j}\left(\rho Y_m v_{m,j}\right) = \rho \dot{\omega}_m \;\; ; \;\; m = 1,2,3.....N
\end{equation}

In-flows to the counterflow enter along the $\chi$ axis. Velocity components and spatial variables in this rotating frame are $u_i = u_{\xi},u_{\chi},w$ and $x_i = \xi,\chi,z$ respectively. The centrifugal acceleration is $a_i = \langle\xi\omega_k^2/4,\chi\omega_k^2/4,0 \rangle$. The quantities $\rho$, $p$, $T$, $h$, $h_m$, $v_m$, $Y_m$, $\lambda$, $c_p$, $c_{p_m}$, $h_{f,m}$, and $\dot{\omega}_m$ are density, pressure, temperature, mixture specific enthalpy, specific enthalpy of species $m$, diffusion velocity of species $m$, mass fraction of species $m$, mixture thermal conductivity, mixture specific heat, specific heat of species $m$, heat of formation of species $m$, and reaction rate of species $m$, respectively. The analysis will use the low-Mach number approximations, i.e. terms on the order of kinetic energy per unit mass are small compared to thermal energy and therefore are negligible. This applies to the last terms on the left and right hand sides of Eq. 4. Additionally, viscous dissipation terms are assumed to be negligible except for cases with large strain rates which are not the subject of this paper and will be analyzed in future work.
\renewcommand{\arraystretch}{2}
\begin{table}[!htb]
\tbl{Normalization quantities}
{\begin{tabular}{|l|c|c|c|c|c|c|} \hline
Variable(s) & $u_j, v_m$ & $t$ & $x_i$ & $\rho$ & $h, h_m, h_{f,m}$ & $p$\\ \hline
Normalization & $\left(\frac{S^*\mu_{\infty}}{\rho_{\infty}}\right)^{1/2}$ & $S^{*^{-1}}$ & $\left(\frac{\mu_{\infty}}{\rho_{\infty}S^*}\right)^{1/2}$ & $\rho_{\infty}$ & $c_{p_{\infty}}T_{\infty}$ & $S^*\mu_{\infty}$ \\ \hline \hline
Variable(s) & $\dot{\omega}_m$ & $\mu$ & $\lambda$ & $c_p, c_{p_m}$ & $\omega_k$ & \\ \hline
Normalization & $S^*$ & $\mu_{\infty}$ & $\mu_{\infty}c_{p_{\infty}}$ & $c_{p_{\infty}}$ & $S^*$ & \\ \hline
\end{tabular}}
\label{table:1}
\end{table}

While the boundary conditions are presented for the similar form in the following paragraphs, we will briefly discuss what they would be for Eqs. (1-5). The energy and species conservation equations use Dirichlet boundary conditions for temperature and mass fractions and are the same in both the Newtonian and non-Newtonian reference frames. The velocities use Neumann boundary conditions to specify imposed strain rates in the three coordinate directions. In this rotating frame, the effect of the vorticity in the original reference frame now appears through a centrifugal acceleration.

The non-dimensional forms of the governing equations remain identical to Eqs. (1-5) when proper normalization factors are chosen. See Table \ref{table:1} for details. Note, dimensional normal strain rates imposed on the $\xi$ and $z$ axes are respectively $S_1^*$ and $S_2^*$ and are each normalized by the principal strain rate at $\infty$ along the $\chi$ axis $S^* \equiv S_1^* + S_2^*$ forming non-dimensional strain rates $S_1$ and $S_2$. Total specific enthalpy is defined as $h_m^0 \equiv h_m + h_{f,m}$ and is normalized according to Table \ref{table:1}. The similarity solution derived by Sirignano is used to reduce the 3D governing equations to a set of ODEs employing the density-weighted Illingworth transformation $\eta \equiv$ \(\int_{0}^{\chi} \rho(\chi^{'}) \,d\chi^{'}\). In the similar form, $u_{\xi} = S_1\xi(df_1/d\eta)$, $w = S_2z(df_2/d\eta)$, and $u_{\chi} = -f/\rho$ with $f \equiv S_1f_1 + S_2f_2$. The several mixed second derivatives of pressure are zero, thereby allowing the simplification of the momentum equations for the $\xi$ and $z$ directions and the formation of the similar form of those equations. The final forms of the governing equations with boundary conditions are presented below. Note, the centrifugal effect is incorporated in the last term of Eq. (6) and in the corresponding boundary condition at $\eta = -\infty$ in Eq. (12). The physical problem remains three-dimensional in space. The reduction to ODEs in one spatial variation follows analytical representation of velocity variation in the other two spatial coordinates.
\begin{equation}
    \rho \mu f_1{'''} + f_1{''}(\rho \mu){'} + f f_1{''} + S_1\left(\frac{1}{\rho} - (f_1{'})^2\right) + \frac{\omega_{k}^{2}}{4S_1}\left(1 - \frac{1}{\rho}\right) = 0
\end{equation}

\begin{equation}
    \rho \mu f_2{'''} + f_2{''}(\rho \mu){'} + f f_2{''} + S_2\left(\frac{1}{\rho} - (f_2{'})^2\right) = 0
\end{equation}

\begin{equation}
   f T^{'} + \frac{1}{c_p}(\rho\lambda T^{'})^{'} - \frac{1}{c_p}T^{'}\sum_{m=1}^N J_m c_{p_m} - \frac{1}{c_p}\sum_{m=1}^N h^0_m\dot{\omega}_m = 0 \;\; ; \;\; m = 1,2,3.....N
\end{equation}

\begin{equation}
    f Y_m^{'} - J_m^{'} + \dot{\omega}_m = 0 \;\; ; \;\; m = 1,2,3.....N
\end{equation}
Definitions and conversions are:
\begin{equation}
    J_m \equiv \rho Y_m v_{m,\eta} = \frac{\rho^2 W_m}{\overline{W}_{mix}^2} \sum_{j\neq m}^N W_j D_{mj} X_j' \;\; ; \;\; v_{m,\eta} = \frac{\rho W_m}{\overline{W}_{mix}^2 Y_m} \sum_{j\neq m}^N W_j D_{mj} X_j' \;\; ; \;\; m = 1,2,3.....N
\end{equation}
\begin{equation}
    X_m = \frac{\overline{W}_{mix}}{W_m}Y_m \;\; ; \;\; \overline{W}_{mix} \equiv \sum_{n=1}^N X_n W_n \;\; ; \;\; m = 1,2,3.....N
\end{equation}
Boundary conditions:
\begin{equation}
    f_1^{'}(\infty) = 1 \;\; ; \;\;  f_1^{'}(-\infty) = \sqrt{\frac{1}{\rho_{-\infty}} + \left(\frac{\omega_k}{2S_1}\right)^2\left(1 - \frac{1}{\rho_{-\infty}}\right)} \;\; ; \;\; f_1(0) = 0
\end{equation}
\begin{equation}
    f_2^{'}(\infty) = 1 \;\; ; \;\;  f_2^{'}(-\infty) = \frac{1}{\sqrt{\rho_{-\infty}}} \;\; ; \;\; f_2(0) = 0
\end{equation}

\begin{equation}
    Y_m(\infty) = Y_{m,\infty} \;\; ; \;\; Y_m(-\infty) = Y_{m,-\infty}
\end{equation}

\begin{equation}
    T(\infty) = 1 \;\; ; \;\;  T(-\infty) = T_{-\infty}
\end{equation}

\subsection{Chemical Kinetics}
The inclusion of multi-step chemical kinetics and the associated thermo-physical properties and diffusion formulation was achieved via integration with Cantera \cite{Cantera}, an open-source toolbox for chemical kinetics, thermodynamics, and transport processes. Cantera allows users to load CHEMKIN-style chemical kinetic models and will compute a host of thermodynamic properties, reaction rates, and transport coefficients. Note that we do not use the counterflow analysis from Cantera; rather, we use our three-dimensional analysis cast in the rotating frame of reference. In the present work, we are interested in hydrogen-oxygen-nitrogen flamelets and will use the hydrogen combustion subset of the Foundational Fuel Chemistry Model 1.0 (FFCM-1) developed at Stanford University \cite{ffcm1}. This model considers 25 reactions among the following nine species: \ce{H2}, \ce{O2}, \ce{H}, \ce{O}, \ce{OH}, \ce{H2O}, \ce{HO2}, \ce{H2O2}, and \ce{N2}. Note that $\mathrm{NO_x}$ formation is not considered in the chemical kinetics model.

\subsection{Numerical Methods}
The computational scheme used to obtain the steady-state solutions of the flamelet equations is a hybrid pseudo-time stepping/Newton method, similar to those used in CHEMKIN OPPDIF and Cantera ``difflame'' solvers. An initial solution is guessed and iterated in pseudo-time until the faster Newton method becomes stable and can iterate the solution to convergence. As this paper is primarily concerned with the effects of vorticity near the flammability limit, a steady-state solution is first obtained at a low strain rate to be used as the initial guess for a higher strain rate. The strain rate is sequentially increased in this manner, using the former steady-state solution as the initial guess for a higher strain rate, until the flame extinguishes. This completes computation of the stable (upper) branch of the S-curve. For the non-premixed flamelet, the program then creates a new guess with a lower strain rate and a lower peak temperature and sequentially decreases these parameters to obtain the unstable (middle) branch of the S-curve. The emphasis is placed on the stable branch for the premixed case.

Our system of governing equations has $N + 3$ dependent variables where $N$ is the number of species, i.e. $Y_1, Y_2, Y_3, ... Y_N, T, f_1^{'}, f_2^{'}$ for mass fractions of all species, temperature, derivative of the $\xi$-momentum similarity variable, and derivative of the $z$-momentum similarity variable. The computational domain of the counterflow is broken into a variable mesh of $n$ spatial nodes. The system is iterated via Newton's method \cite{IntroNumericalAnalysis} and the matrix equation is solved using the generalized minimum residual method built into MATLAB \cite{gmres}.

\subsection{Validation with Cantera}
\begin{figure}[!htb]
\centering
\noindent\makebox[\textwidth]{%
\includegraphics[width=1.22\textwidth]{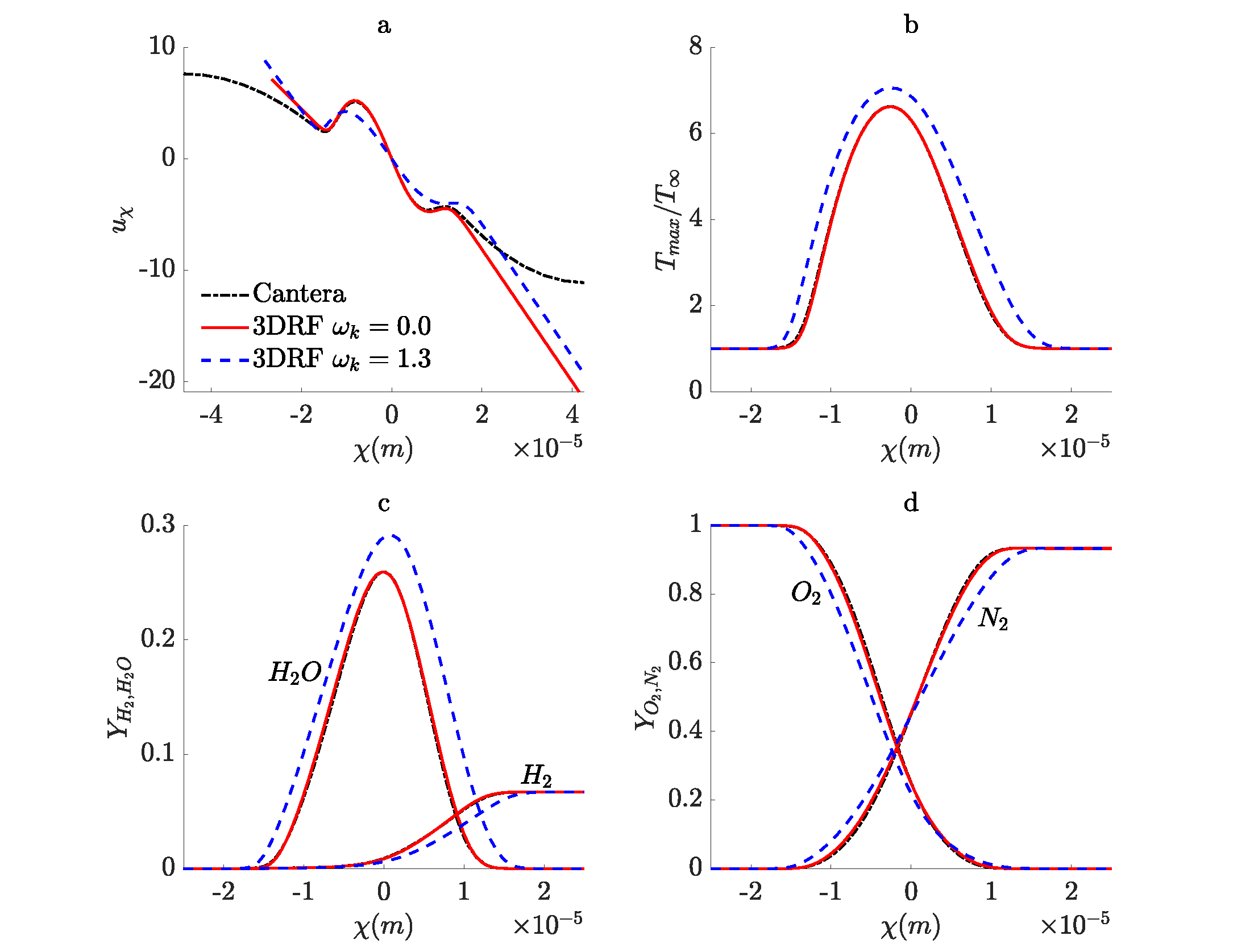}}
\caption{Comparison between the rotational flamelet model and Cantera's default non-premixed flame solver. (a) Axial velocity; (b) Temperature; (c) \ce{H2O} and \ce{H2} mass fractions; (d) \ce{O2} and \ce{N2} mass fractions. The maximum strain rate for matched curves is $S^*_{max}$ = 983,000~1/s and the transverse strain rates in the rotational flamelet model are equal i.e., $S_1 = S_2 = 0.5$ or $S^*_1 = S^*_2 = 491,500$~(1/s), to achieve axisymmetry.}
\label{fig2}
\end{figure}
Computations from the rotational flamelet model were compared with axisymmetric counterflow non-premixed flames from Cantera to validate proper qualitative flame zone behavior. Here, we present the case of a nitrogen-diluted hydrogen-oxygen counterflow non-premixed flame. The oxidizer stream is pure \ce{O2} at 300 K, while the fuel stream is an equimolar mixture of \ce{H2} and \ce{N2} at 300 K. Nitrogen was added to the fuel to shift the location of minimum density into the flame zone. During development, we noted that the centrifugal effect is most prominent when the location of minimum density coincides with the location of maximum (or near-maximum) temperature. Because the centrifugal effect is a core aspect we seek to demonstrate in this study, we imposed the minimum-density/maximum-temperature coincidence to make the centrifugal effect more prominent. Additionally, most fuel-oxidizer combinations in use today have this coincidence; so, we believe this arrangement also helps demonstrate the ubiquity of the centrifugal effect. It was found that approximately $50\%$ $\mathrm{N_2}$ by mole in the fuel stream was sufficient to coincidentally locate minimum density and maximum temperature; so, we picked an equimolar $\mathrm{H_2}$-$\mathrm{N_2}$ fuel for the ease of round numbers. The pressure is set at 10 atmospheres (atm) to achieve a compromise between actual engineering applications which often operate above 10 atm and validation data, both for the chemical kinetics model and laboratory flames, which are predominantly at 1 atm.

A maximum local strain rate $S^*_{max}$ = 983,000~(1/s) is maintained in the flame zone. Equal strain rates along the $\xi$ and $z$ directions, i.e. $S_1 = S_2 = 0.5$ or $S^*_1 = S^*_2 = 491,500$~(1/s), are considered in the rotational flamelet model to match the axisymmetry of Cantera. Axial velocity, temperature, and mass fraction profiles are compared and shown in Fig. \ref{fig2}. There is excellent agreement in the various profiles within the diffusion layer. This is expected because the local maximum strain rates are the same by design (less than $0.16\%$ error). Equivalency of the local strain rate, combined with the central stagnation point in the flame zone requires axial velocity profiles to closely match. Since the diffusion formulation is identical, as is the transport model, it follows that the advective-diffusive-chemical balance is the same, resulting in excellent agreement of scalar curves. Outside the diffusion layer, there are notable differences in velocity resulting from the mathematical framework the two codes are based on. 

Cantera is based on a nozzle counterflow. This means the fuel and oxidizer streams issue from nozzles at a fixed and finite separation distance. Nozzle flow forces the axial velocity gradient to zero at the boundaries as is seen on the black dash-dot curve in Fig. \ref{fig2}\textbf{a}. The incoming streams for the rotational flamelet in the non-Newtonian frame are instead based on a far-field potential counterflow in which the domain extends from $-\infty$ to $\infty$ and the axial strain rate is non-zero at the computational boundary. In these regions outside the flame zone, the model should not be compared with Cantera. The axial strain rate is constant in the far-field, thereby better describing sub-grid turbulent-flow conditions. A viscous layer exists in the interfacial region of the flamelet. In the rotational model, the dimensional strain rate, $S^*$, is defined as the asymptotic axial strain rate in the positive far-field. With regard to extinction behavior, comparisons between the new flamelet model (with or without rotation) and Cantera or CHEMKIN are difficult to make due to different definitions of global strain rates. Nonetheless, the excellent agreement in the flame zones between the rotational flamelet model and the Cantera results demonstrates the thermo-physical validity of the new flamelet model. Also shown in Fig. \ref{fig2} are profiles for $\omega_k = 1.3$. This is to demonstrate the vorticity effect plotted in physical space. A full discussion of vorticity effects is given in Section 3.

In the example shown in Fig. \ref{fig2}, the strain rate for the flamelet is $S^*_{max}$ = 1,000,000~(1/s), the flame-zone dimension is approximately $30~\mu m$, and the viscous-layer thickness is $\delta = (\nu/S^*)^{1/2} \approx 3~\mu m$ with kinematic viscosity $\nu = O(10^{-5}~m^2/s)$. The viscous layer thickness is based on the local $ Re = O(1)$ which makes it the same order as the Kolmogorov scale if the flamelet were embedded in a turbulent field. A situation with these conditions would occur on the smallest scale of the turbulent flow if the integral scale had a shear flow with a velocity variation of $O(100~m/s)$ across a shear layer with transverse dimension $O(10~cm)$, which produces $Re = 10^6$ and a strain rate of $O(1000~1/s)$ on the integral scale. The strain rate, velocity, and length sizes would scale as $Re^{1/2}$, $Re^{-1/4}$, and $Re^{-3/4}$, respectively \cite{Zhu_Antonia_1996,Johnson2024,Hinze_1959,Pope_2000}, as we move from the integral scale to the Kolmogorov scale. One implication is that eddies that are an order of magnitude larger than the Kolmogorov scale could distort and affect the flamelet with unsteady behavior. This is a proper issue for future study. In the following section on non-premixed flames, note that the flame thickness is several times, up to one order of magnitude, larger than the viscous Kolmogorov scale, even near the extinction strain rate. Thus, the flame thickness always exceeds the Kolmogorov scale with our detailed transport treatment. Comparison of Fig. \ref{fig2}\textbf{a} with Fig. \ref{fig2}\textbf{b}, \ref{fig2}\textbf{c}, and \ref{fig2}\textbf{d} indicates that the layer for viscous stress has a thickness comparable with the layer thickness for the scalar variables. Thus, it is suggested that our scaling estimate, which does not neglect the temperature effect on kinematic viscosity, actually underestimates the Kolmogorov scale for reacting flows.

\section{Results}
\subsection{non-premixed flames}
We further examine the non-premixed flame case discussed in the validation section and examine the impacts of vorticity and transverse strain rates. The same mixture composition and conditions are considered, while the axial strain rate is gradually increased to obtain extinction curves. In addition, three combinations of non-dimensional far-field $\xi$ and $z$ strain rates are used, $(S_1 = 0.75$, $S_2 = 0.25)$, $(S_1 = 0.50$, $S_2 = 0.50)$, and $(S_1 = 0.33$, $S_2 = 0.67)$, and three values of non-dimensional vorticity are applied $(\omega_k = 0.0, 1.0, 1.3)$. Note that a single legend applies to all sub-figures.

\begin{figure}[!htb]
\centering
\noindent\makebox[\textwidth]{%
\includegraphics[width=1.22\textwidth]{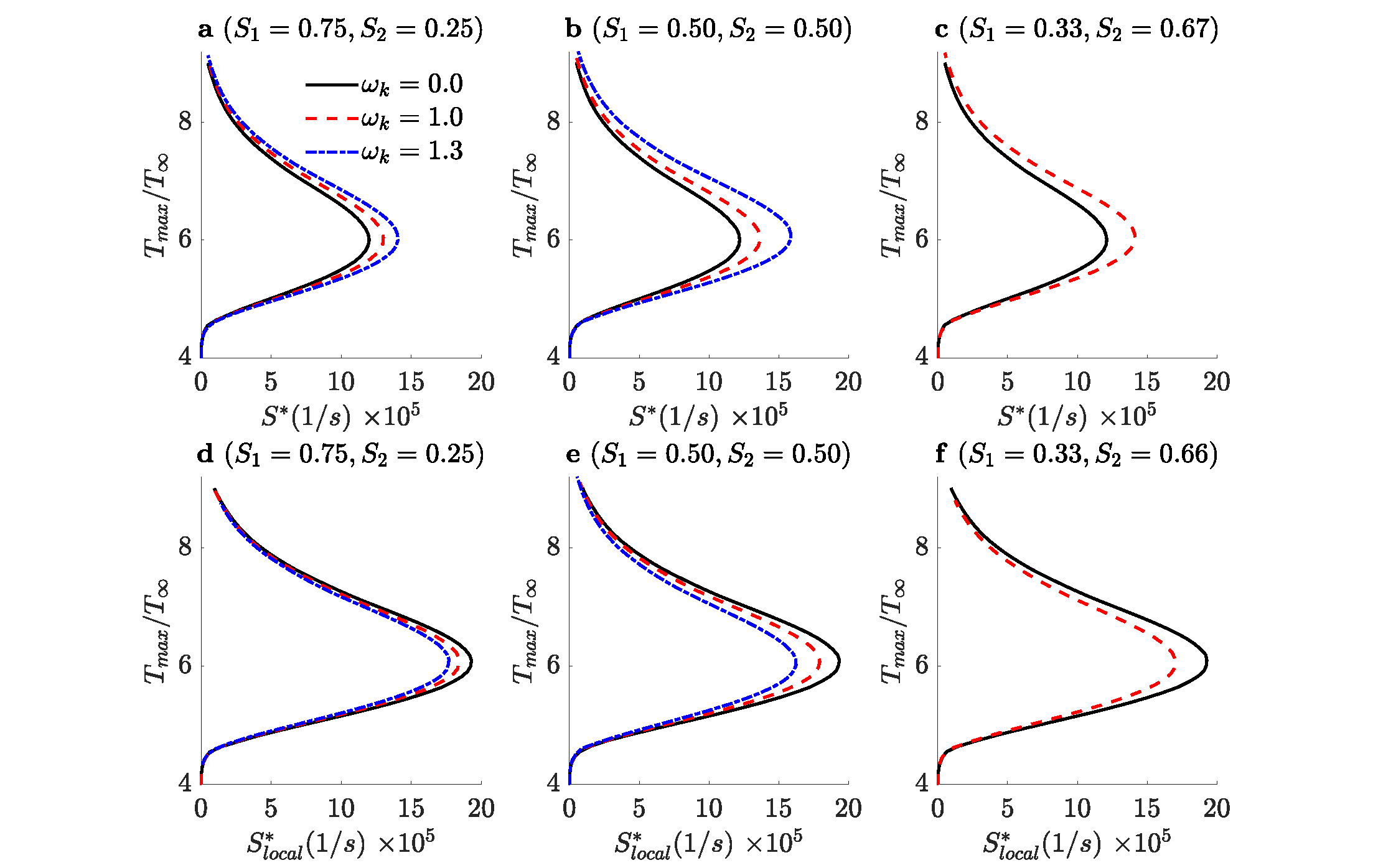}}
\caption{Extinction curves (S-curves) of counterflow non-premixed flames with varying strain rate and vorticity magnitudes. Figure \ref{fig3}\textbf{a}, \ref{fig3}\textbf{b}, and \ref{fig3}\textbf{c} are plotted against ambient strain rate ($S^*$) while Figure \ref{fig3}\textbf{d}, \ref{fig3}\textbf{e}, and \ref{fig3}\textbf{f} are plotted against the local maximum strain rate at the stagnation point ($S^*_{local}$).}
\label{fig3}
\end{figure}
Figure \ref{fig3} shows the extinction curves, also referred to as S-shaped curves or simply S-curves, of the diluted hydrogen-oxygen counterflow non-premixed flame at various combinations of $S_1$, $S_2$, and $\omega_k$. Figure \ref{fig3}\textbf{a}, \ref{fig3}\textbf{b}, and \ref{fig3}\textbf{c} clearly demonstrate that imposing vorticity on the counterflow domain extends the flammability limit in terms of the ambient strain rate, $S^*$. When $S_1 = 0.75$, $S_2 = 0.25$, there is a $17.0\%$ increase in the ambient extinction strain rate between vorticity values of $\omega_k = 0.0$ and $\omega_k = 1.3$. When $S_1 = 0.5$, $S_2 = 0.5$, there is a $30.0\%$ increase in the ambient extinction strain rate for the same vorticity range. When $S_1 = 0.33$, $S_2 = 0.67$, there is a $16.8\%$ increase in ambient extinction strain rate between $\omega_k = 0.0$ and $\omega_k = 1.0$. Note that $\omega_k^*$ changes with $S^*$ due to normalization. Thus, the extinction points for $\omega_k = 1.0$ and $\omega_k = 1.3$ have different dimensional values of vorticity applied. Differences in S-curve behavior based on magnitudes of $S_1$ and $S_2$ are a result of flame stability dependence on mass efflux rates in the $\xi$ and $z$ directions. $S_1$ and $S_2$ are indicative of the percentages of outflow leaving the domain along the $\xi$ and $z$ axes, respectively. However, they only give exactly the outflow percentage if $\omega_k = 0.0$. Note that $\xi$ lies in the plane in which centrifugal force acts on the flow while $z$ is normal to this plane and does not align with a centrifugal force. Thus, when $S_1 > S_2$, a higher percentage of flow exits along the $\xi$-axis and is accelerated by the centrifugal force, reducing residence time and leading to earlier extinction (in terms of $S^*$). Conversely, when $S_1 < S_2$, a higher percentage of flow exits along the $z$-axis and is not accelerated by centrifugal force. In this case, the residence time is not reduced and may increase depending on the relative magnitudes of $S_1$ and $S_2$. Although not shown in the figures, the vorticity effect is consistent at elevated pressure. In these cases, flammability limits are further extended due to the increased kinetic rate caused by increased pressure.

Figure \ref{fig3}\textbf{d}, \ref{fig3}\textbf{e}, and \ref{fig3}\textbf{f} show the same S-curves plotted against the local maximum strain rate at the stagnation point. Opposite behavior, with respect to vorticity, is observed here because the primary function of vorticity is to reduce the strain rate in the flame zone. Thus, at the same maximum temperature, a flame with vorticity has a lower local maximum strain rate. Figure \ref{fig3}\textbf{a}, \ref{fig3}\textbf{b}, and \ref{fig3}\textbf{c} may be thought of as plots of output (ordinate) versus input (abscissa) while Figure \ref{fig3}\textbf{d}, \ref{fig3}\textbf{e}, and \ref{fig3}\textbf{f} may be thought of as output versus output. Analysis of local strain rate is further developed in Figure \ref{fig7} and the accompanying text.

\begin{figure}[!htb]
\centering
\noindent\makebox[\textwidth]{%
\includegraphics[width=1.22\textwidth]{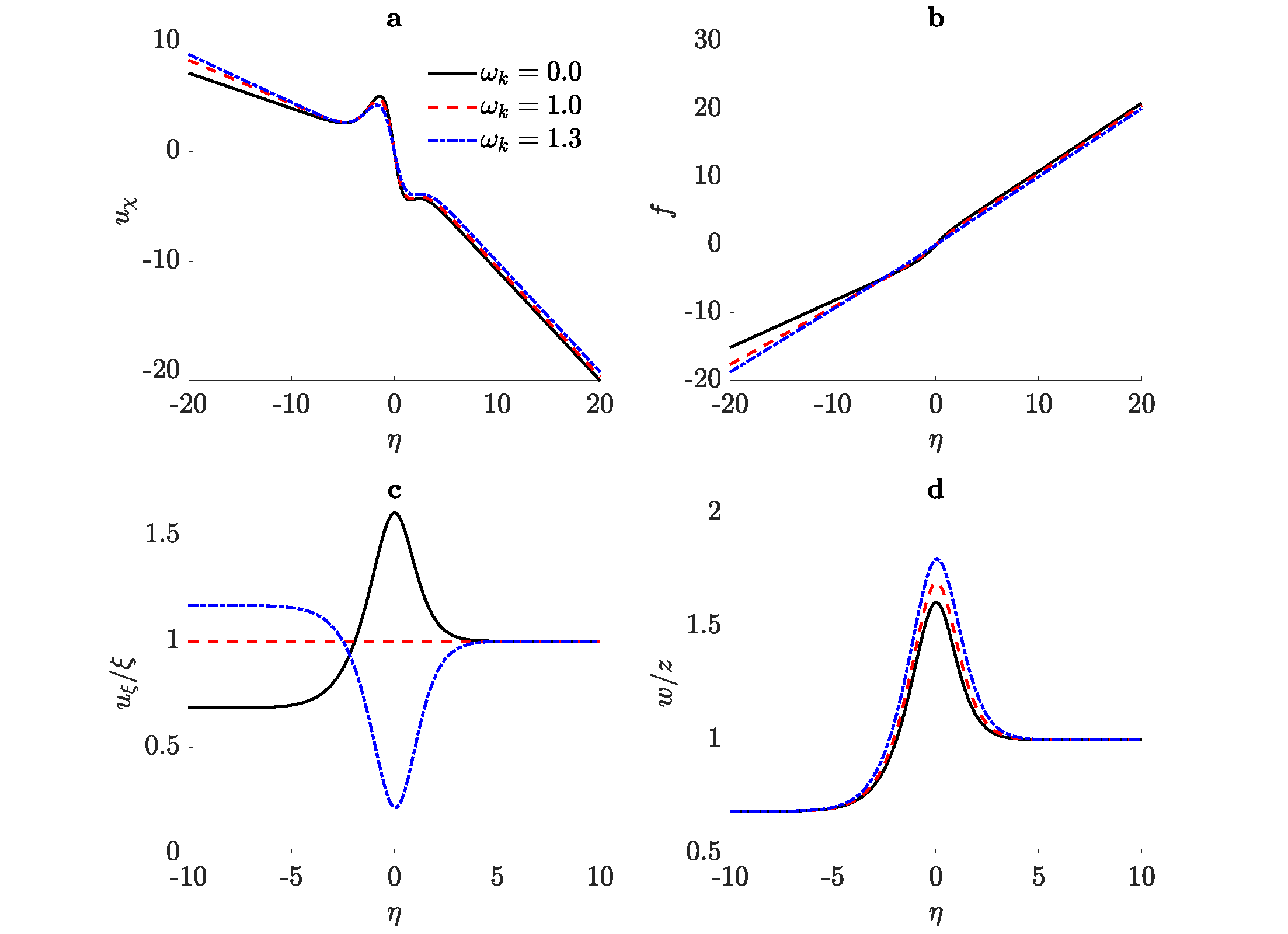}}
\caption{non-premixed flame. (a) Axial (inflow) velocity $u_\chi$; (b) Similarity variable $f$; (c) Transverse outflow velocity normal to vorticity vector divided by length $(u_{\xi}/\xi)$; (d) Transverse outflow velocity parallel to vorticity vector divided by length $(w/z)$. $S_1 = 0.5, S_2 = 0.5$, $S^* = 1,200,000$~1/s.}
\label{fig4}
\end{figure}
Figure \ref{fig4} shows three velocity components as well as the similarity variable $f$, indicative of the axial mass flux, for the symmetric far-field strain rate case $S_1 = 0.5$, $S_2 = 0.5$, and $S^* = 1,200,000$~1/s. However, the flow is only axisymmetric if $\omega_k = 0.0$. All values are non-dimensional. Velocity profiles in the $\xi$ and $z$ directions, i.e. $u_\xi$ and $w$, are divided by their respective spatial variables. Inspection of the slopes of the $u_{\chi}$ velocity profiles (Figure \ref{fig4}\textbf{a}) in the diffusion layer ($-5 \leq \eta \leq 5$) reveals the cause of the increased ambient extinction limits noted in Figure \ref{fig3}. Although the imposed far-field strain rates are the same for all values of vorticity, the strain rate (slope) in the flame zone is reduced as vorticity increases. This essentially allows a flame with vorticity that is highly strained in the far-field to behave like a flame without vorticity that is moderately strained in the far-field. These effects are most prevalent when the location of minimum density coincides with the location of peak temperature. Additional cases not shown here were performed with pure hydrogen in-flowing from the right instead of the equimolar hydrogen-nitrogen mixture of the current discussion. In this case, the location of minimum density lies upstream of the stagnation point, toward the fuel inlet boundary, and well outside the flame zone. Under this condition, the outflow is less severely altered by vorticity due to its higher density, and flammability limits are minimally extended. Cases were also tested using water to dilute the hydrogen stream in place of nitrogen. These cases showed no qualitative difference from the nitrogen case. 

Examination of the four sub-figures together shows that excess oxygen enters the domain along the $\eta$-direction when vorticity is applied. This excess oxygen exits the domain along the $\xi$-axis as shown by the $u_\xi/\xi$ magnitude between $-20 < \eta < 5$. The introduction of vorticity also causes undulating behavior of the $u_\xi$ and $w$-velocity components in the flame zone. Without vorticity, $u_\xi/\xi$ and $w/z$ are equal, indicating symmetric outflow in the $\xi$ and $z$ directions. Furthermore, we see these velocity components are elevated in the flame zone showing the hot (low-density) products of combustion accelerating out of the domain. However, with strong vorticity, the low-density products of combustion are forced toward the stagnation point in the $\xi$ direction under the centrifugal action of vorticity. This leaves only the $z$-direction as a path for combustion products to escape which is why the peaks of the $w/z$ curves increase with increasing vorticity. These effects are compounded when the flamelet exists in an asymmetric strain field which is discussed later on.

\begin{figure}[!htb]
\centering
\noindent\makebox[\textwidth]{%
\includegraphics[width=1.22\textwidth]{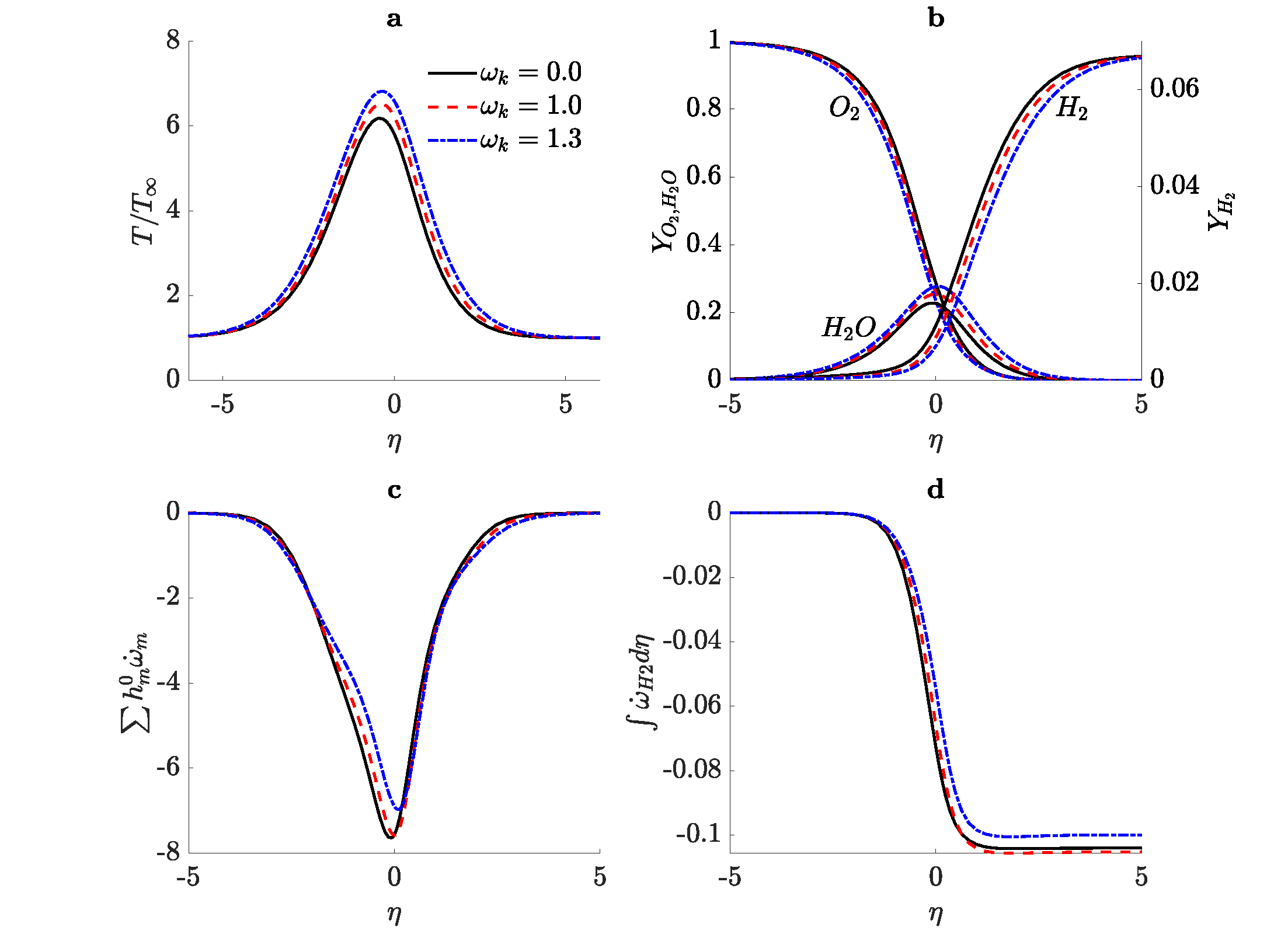}}
\caption{non-premixed flame. (a) Temperature; (b) Major species mass fractions; (c) Heat release rate; (d) Integrated $H_2$ reaction rate (burning rate). $S_1 = 0.5$, $S_2 = 0.5$, $S^* = 1,200,000$~1/s.}
\label{fig5}
\end{figure}
We now consider the effects of vorticity and transverse strain rates on temperature, species, and energy profiles for the case shown in Figure \ref{fig4}. Figure \ref{fig5} shows temperature, major species, heat release rate (HRR), and the integrated \ce{H2} reaction rate profiles for the afore-discussed three cases. Inspection of the temperature profiles shows the direct relationship between peak temperature and vorticity. Next, we examine the species profiles. Unsurprisingly, reactant mass fractions decrease while major product mass fractions increase with vorticity. Essentially, the reduction in strain rate in the flame zone caused by vorticity allows for more complete combustion. Figure \ref{fig5}\textbf{c} and \ref{fig5}\textbf{d} show the decreased heat release rate and burning rate caused by vorticity. This is the result of decreased mass flux through the domain when the centrifugal force acts on the counterflow as previously noted \cite{Sirignano1}.

\begin{figure}[!htb]
\centering
\noindent\makebox[\textwidth]{%
\includegraphics[width=1.22\textwidth]{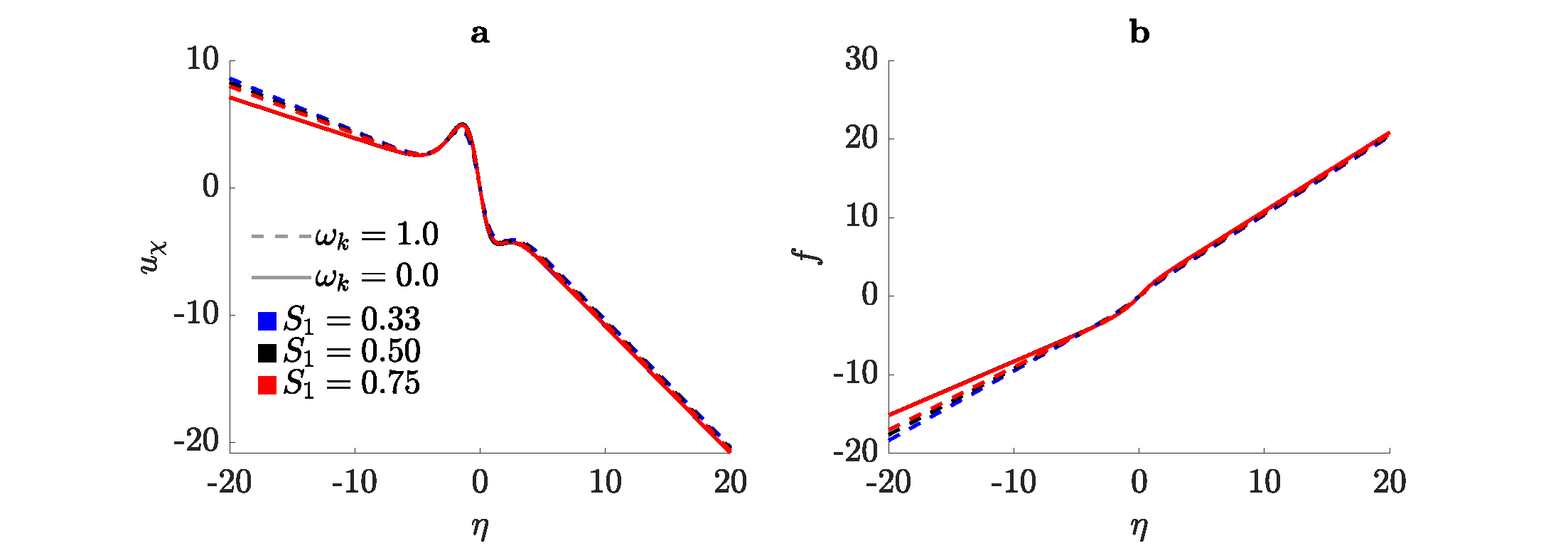}}
\caption{non-premixed flame. (a) Axial (inflow) velocity $u_\chi$; (b) Similarity variable $f$. $S_1 = 0.33, 0.50, 0.75; S_2 = 0.67, 0.50, 0.25$, $\omega_k = 0.0, 1.0$, $S^* = 1,200,000$ $1/s$. The three curves on each sub-figure for $\omega_k = 0.0$ are coincident with the solid red curve.}
\label{fig6}
\end{figure}

\begin{figure}[!htb]
\centering
\noindent\makebox[\textwidth]{%
\includegraphics[width=1.22\textwidth]{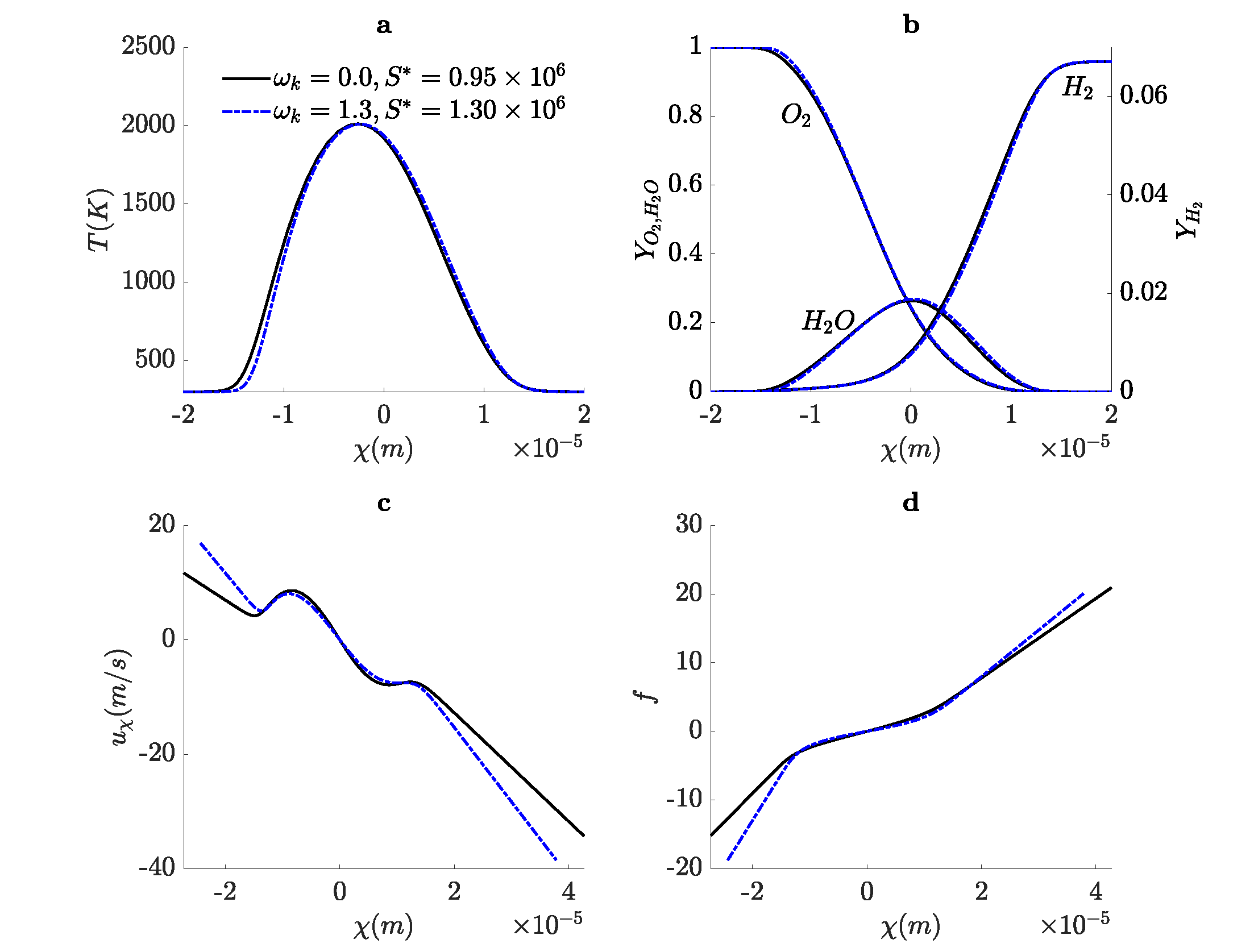}}
\caption{non-premixed flame. (a) Absolute temperature; (b) Major species mass fractions; (c) Dimensional axial velocity $(u_\chi)$; Similarity variable $f$. $S_1 = S_2 = 0.50$, $\omega_k = 0.0, 1.3$, $T_{max} = 2010 K$.}
\label{fig7}
\end{figure}

Besides vorticity, differing strain rates between $\xi$ and $z$ directions can also alter flamelet behaviors compared to equal strain rate scenarios. Figure \ref{fig6} shows the axial velocity $(u_\chi)$ and similarity variable $(f)$ for varying vorticity values $(\omega_k = 0.0, 1.0)$ and transverse strain rates $S_1 = 0.33, 0.50, 0.75$, $S_2 = 0.67, 0.5, 0.25$. Note that for the three cases with $\omega_k = 0.0$ (solid lines), the curves of both the $u_\chi$ and $f$ lie within the plotted line thickness of each other and are thus indistinguishable, i.e. all curves for $\omega_k = 0.0$ are coincident with the solid red line. Certain general trends are unchanged for this asymmetric case, namely the decreased strain rate in the flame zone as well as increased oxygen influx, both a result of imposing vorticity. There is an important difference however. When $S_2 > S_1$, more outflow is allowed in the $z$-direction which is unrestricted by centrifugal force. This allows more oxygen into the domain and reduces the maximum strain rate in the flame zone. Conversely, when $S_2 < S_1$, the opposite trend is observed. 

As is well known, it is the local strain rate in the flame zone, not a distant ambient strain rate that impacts extinction. Figure \ref{fig7} compares various convective and scalar profiles of two flames on the stable branches of Figure \ref{fig3}\textbf{b} having the same maximum temperature to within 0.055$\%$. This comparison is made between the $\omega_k = 0.0$ and $\omega_k = 1.3$ curves. One may expect that all profiles between these cases would locally collapse, based on the assumption that equal maximum temperatures imply equal strain rates in the flame zone. This is not the case, although the local profiles are surely similar. The local maximum strain rates, as seen in Figure \ref{fig7}\textbf{c} differ by approximately 86,000 $1/s$ or 8$\%$ due to the variance in mass efflux caused by vorticity. It follows from the difference in convective gradients that the scalar gradients differ as well, slight though they may be. Now, one may consider small differences in these profiles as superficially unimportant. However, the ability of a flamelet model to predict similar scalar curves from vastly different $S^*$ and $\omega_k$ input parameters is critical to capturing the chaos of turbulent reacting flows. Existing flamelet models fail in this realm because they do not represent both variables, strain rate and vorticity, while mapping resolved flow quantities to sub-grid inputs.

\begin{figure}[!htb]
\centering
\noindent\makebox[\textwidth]{%
\includegraphics[width=1.22\textwidth]{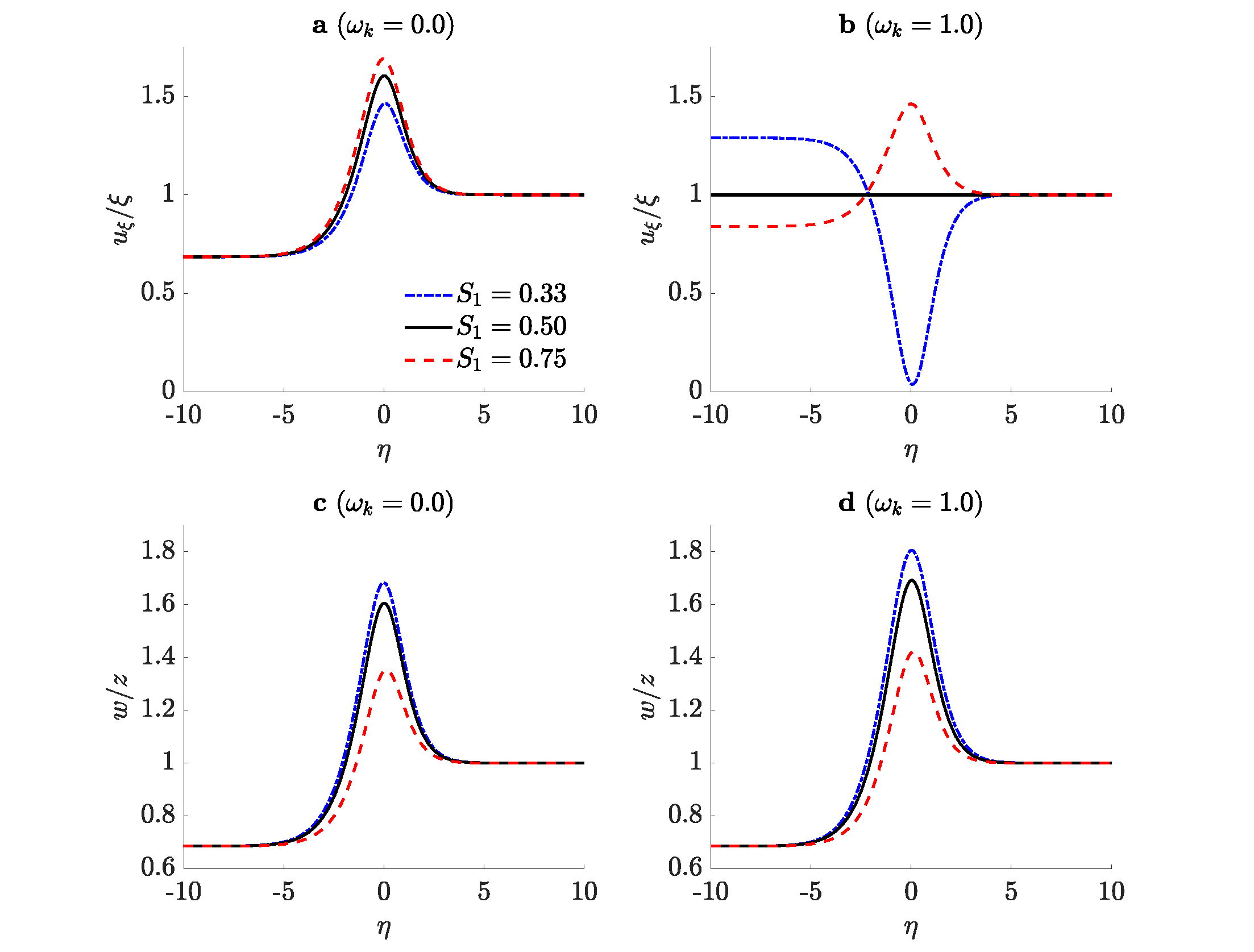}}
\caption{non-premixed flame. (a) $u_{\xi}/\xi$, $\omega_k = 0.0$; (b) $(u_{\xi}/\xi)$, $\omega_k = 1.0$; (c) $(w/z)$, $\omega_k = 0.0$; (d) $(w/z)$, $\omega_k = 1.0$. $S_1 = 0.33, 0.50, 0.75$;  $S_2 = 0.67, 0.50, 0.25$. $S^* = 1,200,000$ $1/s$.}
\label{fig8}
\end{figure}
Figure \ref{fig8} shows the velocity components in the $\xi$ and $z$ directions for $\omega_k = 0.0, 1.0$ and $S_1 = 0.33, 0.50,$ $0.75$; $S_2 = 0.67, 0.5, 0.25$. First, compare the $u_\xi/\xi$ profiles between Figure \ref{fig8}\textbf{a} and \ref{fig8}\textbf{b} for $S_1 = 0.50$, $S_2 = 0.50$. In Figure \ref{fig8}\textbf{b} a vorticity value of $\omega = 1.0$ (solid black) produces a straight line due to the elimination of any density effect in the $\xi$-momentum equation, Equation (5), and the boundary condition given by Equation (11). When $\omega \neq 1.0$ and/or $S_1 \neq 0.5$, the boundary conditions at $-\infty$ and $+\infty$ are not equal, and a straight line does not occur. 

Further comparison of the $u_\xi/\xi$ profiles in Figure \ref{fig8} for the case where $S_1 = 0.75, S_2 = 0.25$ (dash red) shows a conflict of influence in the flame zone between $S_1$, $S_2$, and $\omega_k$. When $S_1 > S_2$, mass efflux is decreased (limited) in the $z$-direction and increased in the $\xi$-direction. Because efflux in the $\xi$-$\eta$ plane feels the centrifugal effect of vorticity while efflux in the $z$-direction does not, the centrifugal effect attempts to expel the expanding low-density products of combustion along the $z$-direction, but the low value of $S_2$ does not permit this. Instead, due to the bias toward $\xi$-efflux from $S_1 > S_2$, $u_\xi/\xi$ is forced to remain positive in the flame zone. Comparing this case $(S_1 > S_2)$ to the case where $(S_1 < S_2)$ (dash-dot blue), we see an alignment between the preferred efflux direction and vorticity magnitude. When vorticity is applied, efflux prefers to flow in $z$, and is restricted in $\xi$, which is why $u_\xi/\xi$ becomes negative and the maximum of $w/z$ is much higher for $S_1 < S_2$ than for $S_1 > S_2$. Overall, this results in a lower strain rate in the flame zone and increased residence time which explains why extinction points in Figure \ref{fig3}\textbf{c} occur at a higher strain rate than the corresponding points in Figure \ref{fig3}\textbf{a} and \ref{fig3}\textbf{b}.

\begin{figure}[!htb]
\centering
\noindent\makebox[\textwidth]{%
\includegraphics[width=1.22\textwidth]{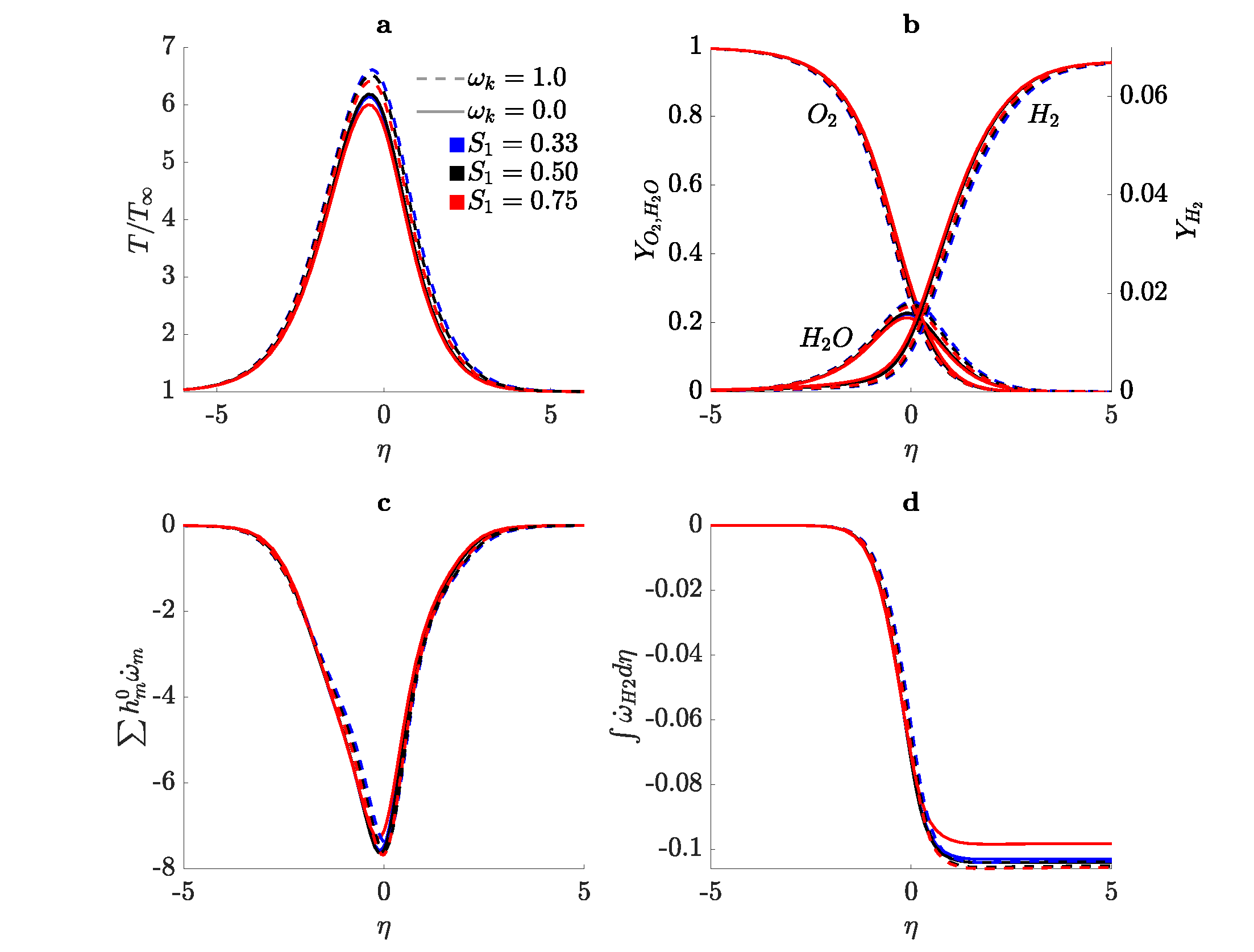}}
\caption{non-premixed flame. (a) Temperature; (b) Major species mass fractions; (c) Heat release rate; (d) Integrated $H_2$ reaction rate (burning rate). $S_1 = 0.33, 0.50, 0.75$;  $S_2 = 0.67, 0.50, 0.25$. $\omega_k = 0.0, 1.0$, $S^* = 1,200,000$ $1/s$.}
\label{fig9}
\end{figure}
Figure \ref{fig9} shows temperature, species, and energy profiles for the symmetric and asymmetric strain rate cases at $\omega_k = 0.0, 1.0$. Many curves are close to one another because axial convection differs substantially only in the negative far-field where there are no scalar gradients. Still, the deviations of strain rate and residence time arising from differing choices of $S_1$, $S_2$, and $\omega_k$, are reflected in the scalar profiles. When $S_1 < S_2$ and $\omega_k = 1.0$, we see the minimum strain rate in the flame zone and the highest residence time, temperature, and product mass fraction. Interestingly, this case also produces the lowest HRR magnitude relative to the other rotational cases. When $S_1 > S_2$ and $\omega_k = 0.0$, we see the maximum strain rate in the flame zone and a reduction in residence time. This case produces the lowest temperature, product mass fraction, and HRR magnitude of the irrotational cases. Profiles of dependent variables on the unstable branch are qualitatively similar to those on the stable branch. While vorticity increases the maximum temperature on the stable branch, it decreases the maximum temperature on the unstable branch. Limits also exist on the maximum magnitudes of $\omega_k$ and $S_2$ that may be applied to the model. When either of these parameters exceed a critical value, the centrifugal momentum overpowers the in-flow momentum and a reversal of flow occurs near the stagnation point. When this happens a pure counterflow no longer exists, instead the mass flux resembles the superposition of a counterflow and a source flow.

The counterflow analysis is applied only at the smallest (i.e. Kolmogorov) length scales to avoid fluctuations from smaller eddies appearing in the layer. Smaller scales could not be considered small perturbations since they would be expected to have larger strain rates, vorticity, and scalar dissipation rates. Also, they would have shorter time scales causing problems with the quasi-steady assumption. The common gases for combustion have kinematic viscosity $\nu$, thermal diffusivity $\alpha$, and mass diffusivity $D$, all very close in magnitudes. Therefore, we expect the viscous layer near the counterflow interface and the non-premixed flame thickness to have similar size. The diffusion layer centers very close to the interface for the non-premixed flame.

\subsection{Premixed Flames}
Here, we present the case of a stoichiometric hydrogen-air premixed twin flame. A combustible mixture of stoichiometric hydrogen and oxygen at 300 K enters the domain at $\eta = \pm \infty$. Due to the symmetry of the problem, only the positive half of the domain is calculated. The pressure is set at 10~atm. In the following figures and discussion, three combinations of $\xi$ and $z$ strain rates are used, $S_1 = 0.47, 0.50, 0.75$;  $S_2 = 0.53, 0.50, 0.25$; and two values of vorticity are applied, $\omega_k = 0.0, 1.0$.

\begin{figure}[!htb]
\centering
\noindent\makebox[\textwidth]{%
\includegraphics[width=1.22\textwidth]{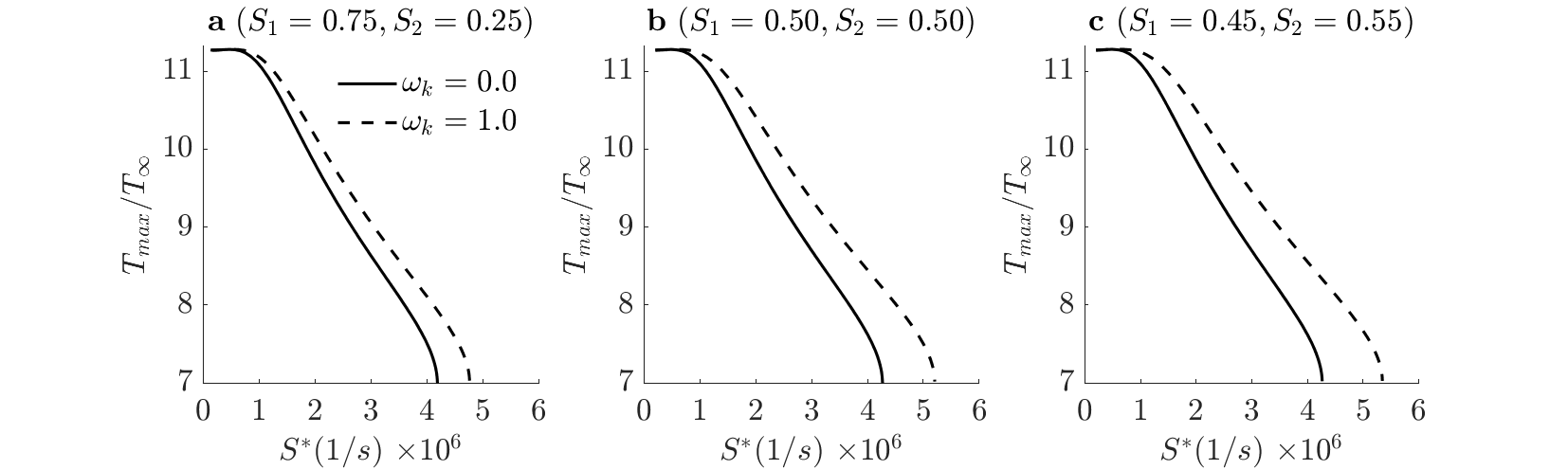}}
\caption{Extinction curves (S-curves) of premixed twin flames with varying strain rate and vorticity magnitudes plotted against ambient strain rate ($S^*$).}
\label{fig10}
\end{figure}
Vorticity has a similar effect on the premixed twin flames as evidenced by Fig. \ref{fig10}. In all variations of transverse strain rate, the S-curve with vorticity extinguishes at a higher ambient strain rate. The preference of cases with larger $S_2$ values extinguishing at higher ambient strain rates is also maintained. We present here only the stable branches of S-curves for the premixed twin flames due to convergence difficulty on the unstable branch. Calculation of both branches will be a goal of future work.

\begin{figure}[!htb]
\centering
\noindent\makebox[\textwidth]{%
\includegraphics[width=1.22\textwidth]{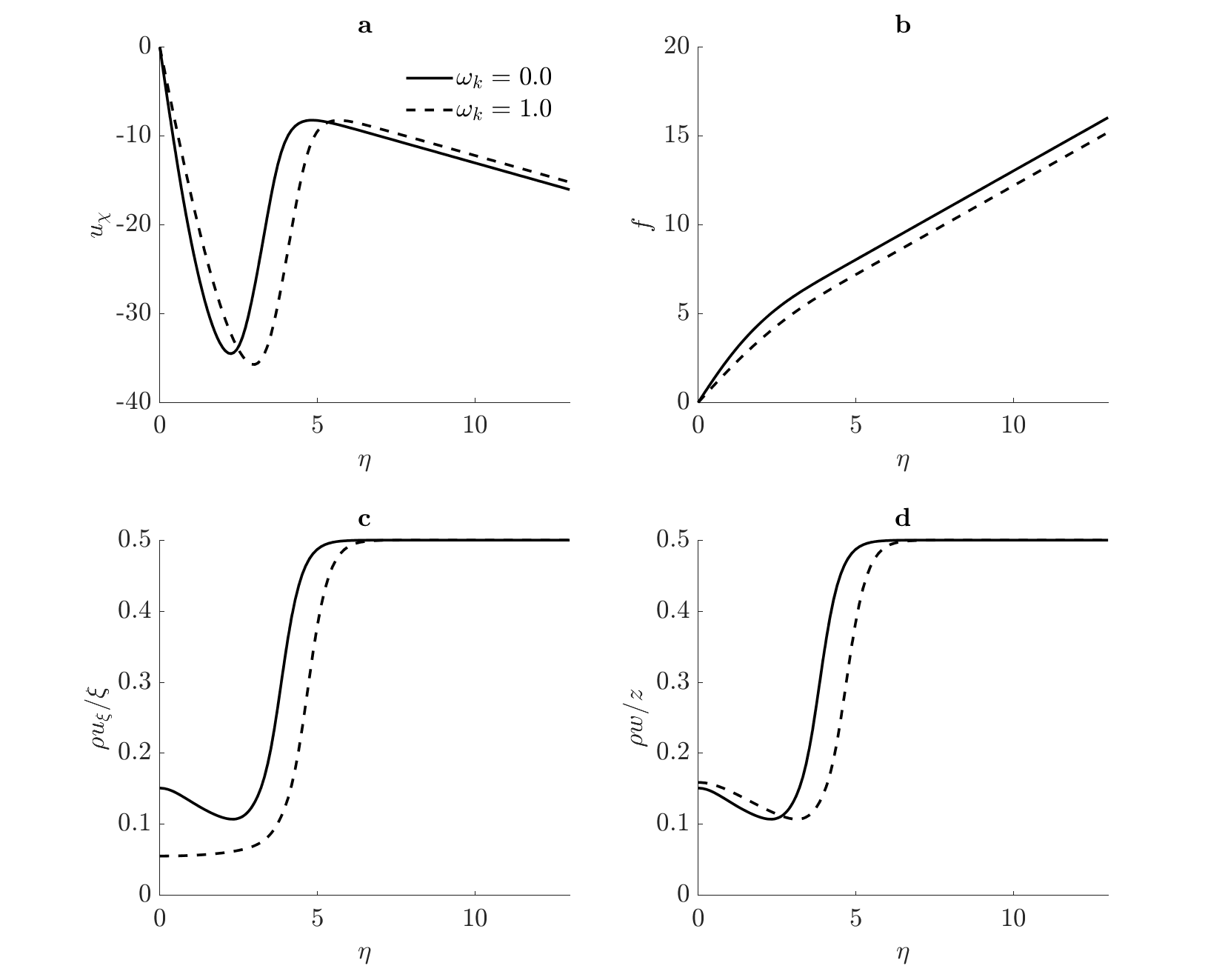}}
\caption{Premixed flame. (a) Axial (inflow) velocity $u_\chi$; (b) Similarity variable $f$; (c) Transverse mass efflux normal to vorticity vector divided by length $(\rho u_{\xi}/\xi)$; (d) Transverse mass efflux parallel to vorticity vector divided by length $(\rho w/z)$. $S_1 = 0.5$; $S_2 = 0.5$; $\omega_k = 0.0, 1.0$; $S^* = 1,000,000$ $1/s$.}
\label{fig11}
\end{figure}
Figure \ref{fig11} shows axial velocity $u_\chi$, the similarity variable $f$, and mass effluxes in the $\xi$ and $z$ directions, $\rho u_{\xi}/\xi$ and $\rho w/z$, respectively, for the symmetric far-field strain rate case $S_1 = 0.5$, $S_2 = 0.5$ at $\omega_k = 0.0, 1.0$. These profiles are taken at an intermediate ambient strain rate of $S^* = 1,000,000$~1/s. All values are non-dimensional. Efflux profiles, in the $\xi$ and $z$ directions, are divided by their respective spatial variables. Figure \ref{fig11}\textbf{a} and \ref{fig11}\textbf{b} show a reduction in boundary velocity and mass flux due to the centrifugal effect. Figure \ref{fig10}\textbf{c} shows a decrease in efflux component $\rho u_{\xi}/\xi$ when vorticity is applied. This occurs to satisfy mass conservation as a higher percentage of fluid exits in the $z$ direction when $\omega_k = 1.0$ according to Fig. \ref{fig11}\textbf{d}. The rightward shift in all sub-figures for $\omega_k = 1.0$ is a consequence of the centrifugal effect reducing inflow velocity.

\begin{figure}[!htb]
\centering
\noindent\makebox[\textwidth]{%
\includegraphics[width=1.22\textwidth]{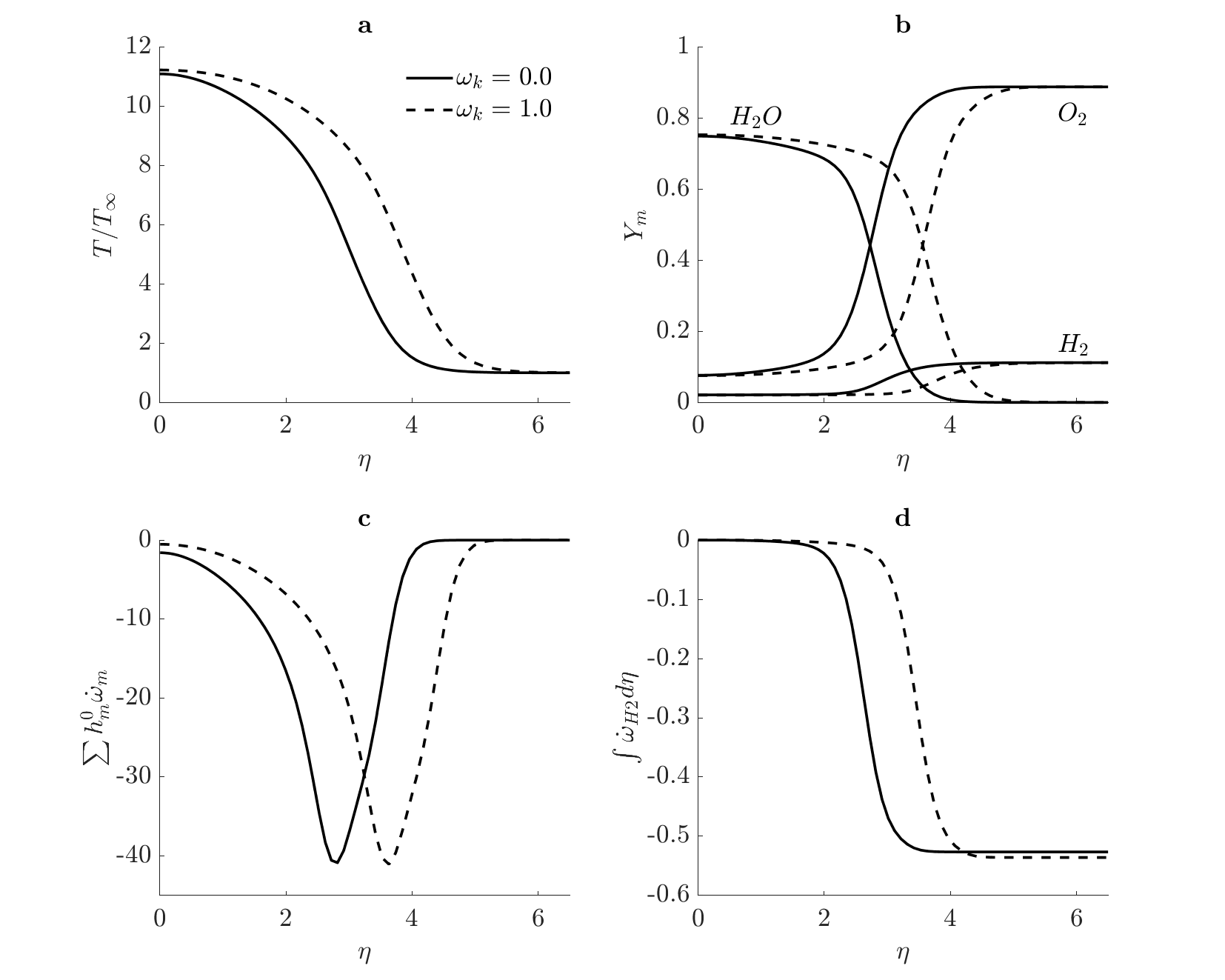}}
\caption{Premixed flame. (a) Temperature; (b) Major species mass fractions; (c) Heat release rate; (d) Integrated $H_2$ reaction rate (burning rate). $S_1 = 0.5$; $S_2 = 0.5$; $\omega_k = 0.0, 1.0$; $S^* = 1,000,000$ $1/s$.}
\label{fig12}
\end{figure}
As the premixed flame does not have a unique peak temperature until the gradient region produced by the flame meets the stagnation point, it is beneficial to discuss the flame standoff distance in addition to S-curves. We define the standoff distance as the distance between the flame zone diffusion layer and the stagnation point, which decreases predictably with $S^*$. Figure \ref{fig12}\textbf{a} shows both an increased peak temperature and increased standoff distance due to the centrifugal effect. Contrary to non-premixed flames which have enhanced combustion via increased residence time, premixed flames are much less dependent on residence time because diffusion is unnecessary to create a flammable mixture. Instead, the effect of vorticity is to reduce the velocity at a given axial location which allows the flame speed, relatively unchanged by vorticity, to move the flame to the right. This results in increased peak temperature because the scalar gradients are removed from the interface, i.e., the flame thickness increases. A comparison can be drawn here with the work of Kolla and Swaminathan \cite{Swaminathan} which examined strained premixed flames. They found that the flame brush thickness increases with the magnitude of the turbulent fluctuating velocity component normalized by unstrained laminar flame speed. Our results are in qualitative agreement as their parameterizing quantity can be likened to our non-dimensional vorticity, $\omega_k = \omega_k^*/S^*$. Our results show modifications of velocity via differing strain rates which was also reported by Knudsen et al. \cite{KnudsenKollaHawkesPitsch}, although that work considered only normal strain rate without vorticity. The species profiles in Figure \ref{fig12}\textbf{b}, heat release rate in \ref{fig12}\textbf{c}, and integrated \ce{H2} reaction rate in \ref{fig12}\textbf{d}, reflect this shift of the diffusion layer. We expect that the temperature of the inflow will affect the flame character near the extinction limit.

\begin{figure}[!htb]
\centering
\noindent\makebox[\textwidth]{%
\includegraphics[width=1.22\textwidth]{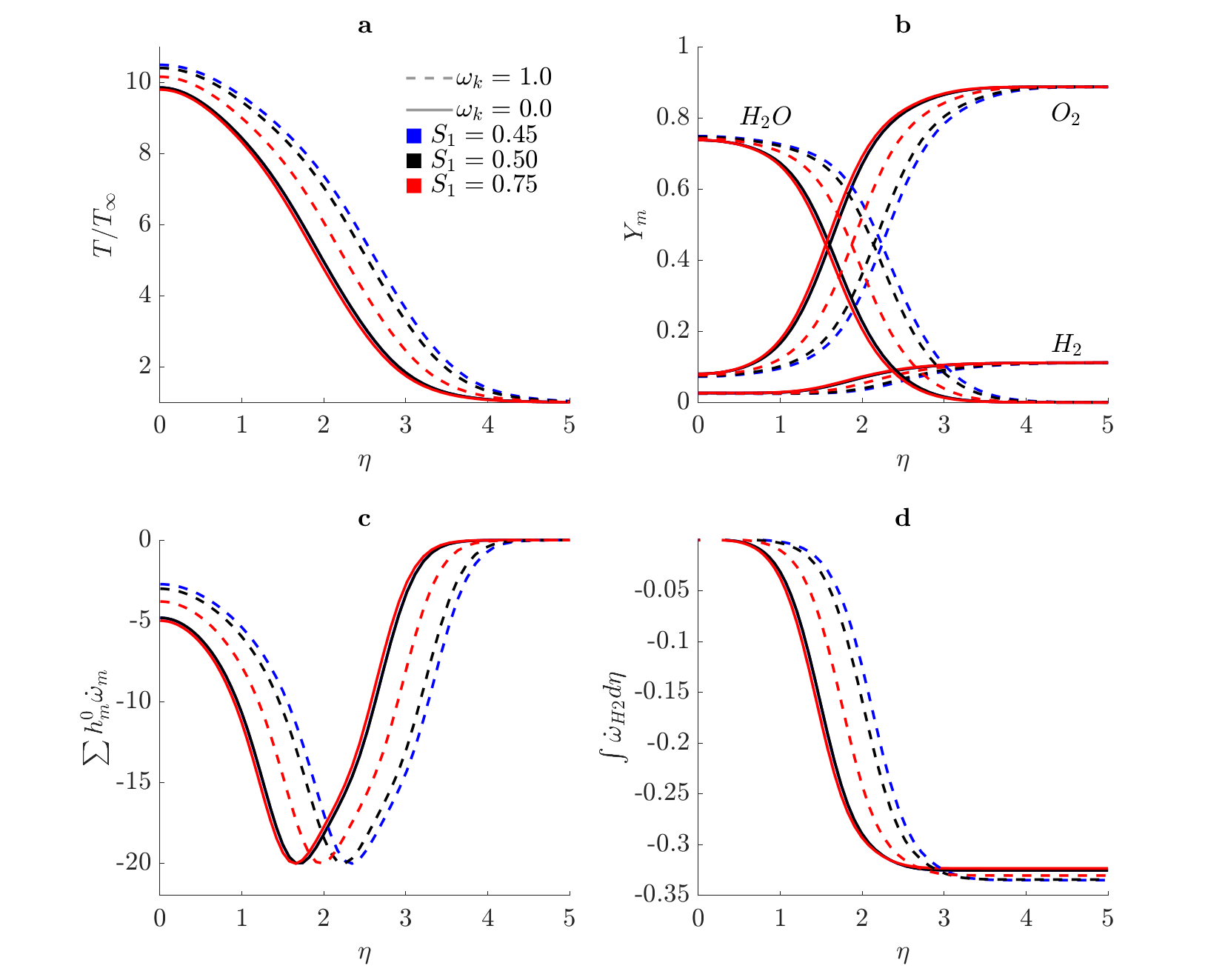}}
\caption{Premixed flame. (a) Temperature; (b) Major species mass fractions; (c) Heat release rate; (d) Integrated $H_2$ reaction rate (burning rate). $S_1 = 0.44, 0.50, 0.75$;  $S_2 = 0.55, 0.50, 0.25$; $\omega_k = 0.0, 1.0$; $S^* = 2,000,000$ $1/s$.}
\label{fig13}
\end{figure}
\begin{figure}[!htb]
\centering
\noindent\makebox[\textwidth]{%
\includegraphics[width=1.22\textwidth]{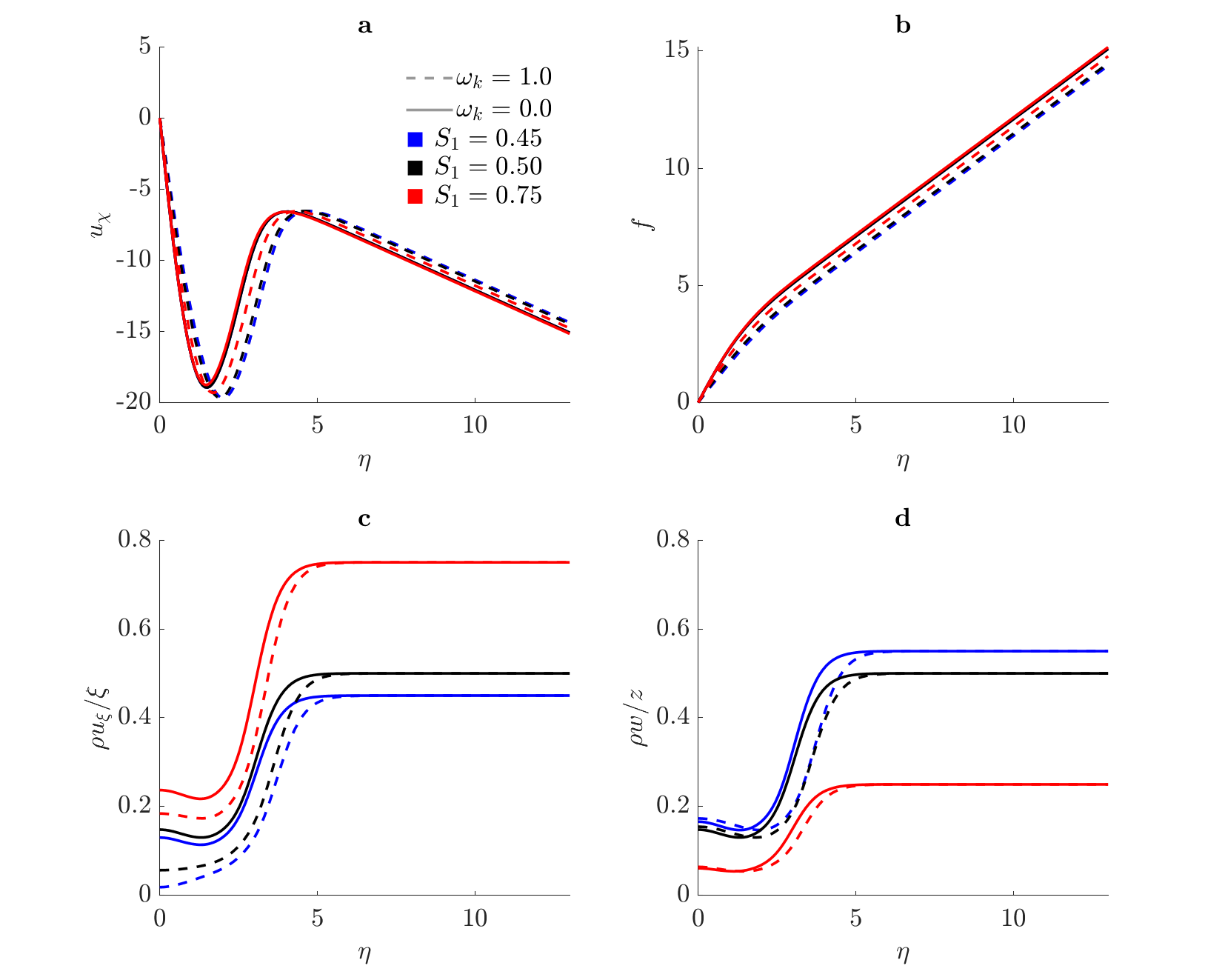}}
\caption{Premixed flame. (a) Axial (inflow) velocity $u_\chi$; (b) Similarity variable $f$; (c) Transverse mass efflux normal to vorticity vector divided by length $(\rho u_{\xi}/\xi)$; (d) Transverse mass efflux parallel to vorticity vector divided by length $(\rho w/z)$. $S_1 = 0.45, 0.50, 0.75$;  $S_2 = 0.55, 0.50, 0.25$; $\omega_k = 0.0, 1.0$; $S^* = 2,000,000$ $1/s$.}
\label{fig14}
\end{figure}

Figures \ref{fig13} and \ref{fig14} explore the same premixed twin flame at a higher strain rate of $S^* = 2,000,000$~1/s with varying $S_1$, $S_2$, and $\omega_k$ combinations. Note, within these figures, some of the $\omega_k = 0.0$ curves are coincident and difficult to distinguish. Figure \ref{fig13} shows that premixed flames are similarly affected by vorticity and $S_2 > S_1$ combination as compared to non-premixed flames. Increasing vorticity increases peak temperature and flame standoff distance. We also notice that when $\omega_k = 0.0$, the case of $S_1 = 0.5, S_2 = 0.5$ produces the highest temperature, as was the case with non-premixed flames. Figure \ref{fig14} shows the velocity and mass flux profiles. These profiles differ from non-premixed flames due to boundary-value densities. At the left boundary, non-premixed flames had increased mass flux when $\omega_k = 1.0$ due to the higher density there; however, the premixed flame has a lower density there and thus a lower mass flux. Note, these density differences also impose a limit on the minimum magnitude of $S_1$.

The Karlovitz number Ka is often used to describe strained premixed flames and to compare flame scales to Kolmogorov scales. It can be written as the square of the ratio of premixed flame thickness to the Kolmogorov length:  $Ka = (\delta_{PF}/ \delta)^2 $ where the flame thickness can be assumed to be dominated by the pre-heat layer so that $\delta_{PF} = D/S_L$ and $S_L$ is the premixed  laminar flame speed. The mass diffusivity $D$ equates with both the kinematic viscosity $\nu$ and the thermal diffusivity $\alpha$.  For the counterflow at the highest strain rate (i.e. Kolmogorov value), the Kolmogorov viscous-layer thickness $\delta = \sqrt{D/S^*}$ where $S^*$ is the highest strain rate in the turbulent flow field; i.e. the value for the Kolmogorov scale in a shear flow is $O(Re^{1/2})$  larger than strain rates on the resolved scale in LES or RANS.  A length infrequently discussed in premixed flamelet studies is the distance $L$  of the flame from the interface in the counterflow. It follows that $LS^* = S_L$ for a quasi-steady behavior with a stationary or near-stationary distance between the flame and interface. Consequently, $Ka = (\delta / L)^2$. So, for $Ka >> 1$, the premixed flame is thick compared to the viscous layer but its center lies close to the counterflow interface. On the other hand, if $Ka << 1$, the premixed flame is thin compared to the viscous layer but the distance from the interface and the viscous layer is large. This is known to be the domain for corrugated or wrinkled laminar flames which occur with moderate levels of turbulence. Our work focuses on high strain rates with a higher degree of turbulence (i.e. higher resolved-scale Reynolds number) where the premixed flame is held in the viscous layer of the strained flow, i.e. cases with $Ka \geq 1$.

\subsection{Partially-premixed Flames}
The rotational flamelet model also has the capability to compute partially-premixed flames. Below we present a flame with a slightly lean mixture inflowing from the left (1.666 \ce{H2}, 1.000 \ce{O2}, 3.760 \ce{N2} in moles) and a rich mixture inflowing from the right (4.950 \ce{H2}, 1.000 \ce{O2}, 3.760 \ce{N2} in moles). Both streams are at 300 Kelvin. At a low strain rate, these boundary conditions produce a counterflow with two major reaction zones (premixed) and one minor reaction zone (diffusion), see Fig. \ref{fig16}\textbf{c}. A lean premixed flame exists around $\eta = -4$, a non-premixed flame exists around $\eta = -1.9$, and a rich premixed flames exists around $\eta = +4$. This is an example of the established triple flame \cite{Kioni1993,Li2020,Plessing1998,NguyenSirignano2018} which can occur post-extinction, once the intensity of turbulent straining decreases. These mixtures were chosen intentionally to produce a triple flame and were found by trial and observation. In the following results, one combination of transverse strain rates is used, $S_1 = S_2 = 0.5$ and two values of vorticity are applied, $\omega_k = 0.0, 1.0$. S-curves, scalar profiles, and velocity profiles are presented below.

\begin{figure}[!htb]
\centering
\noindent\makebox[\textwidth]{%
\includegraphics[width=0.5\textwidth]{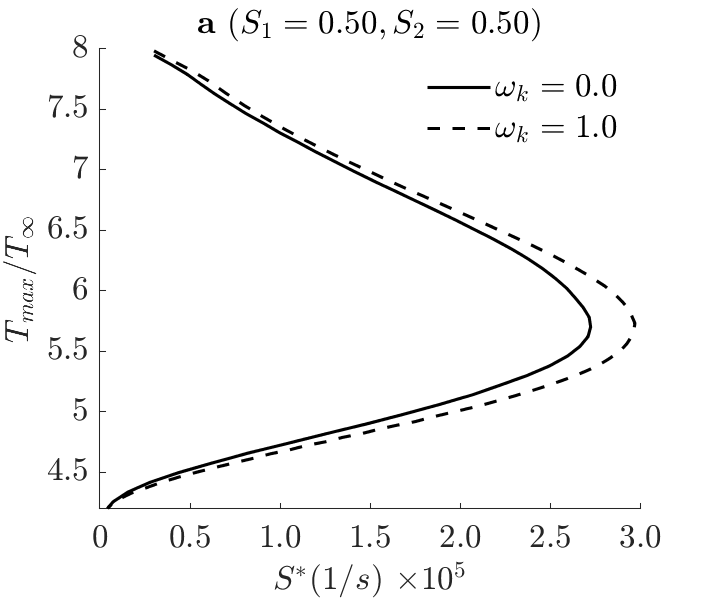}}
\caption{Extinction curves (S-curves) of partially-premixed flames with varying vorticity magnitudes plotted against ambient strain rate ($S^*$).}
\label{fig15}
\end{figure}

\begin{figure}[!htb]
\centering
\noindent\makebox[\textwidth]{%
\includegraphics[width=1.22\textwidth]{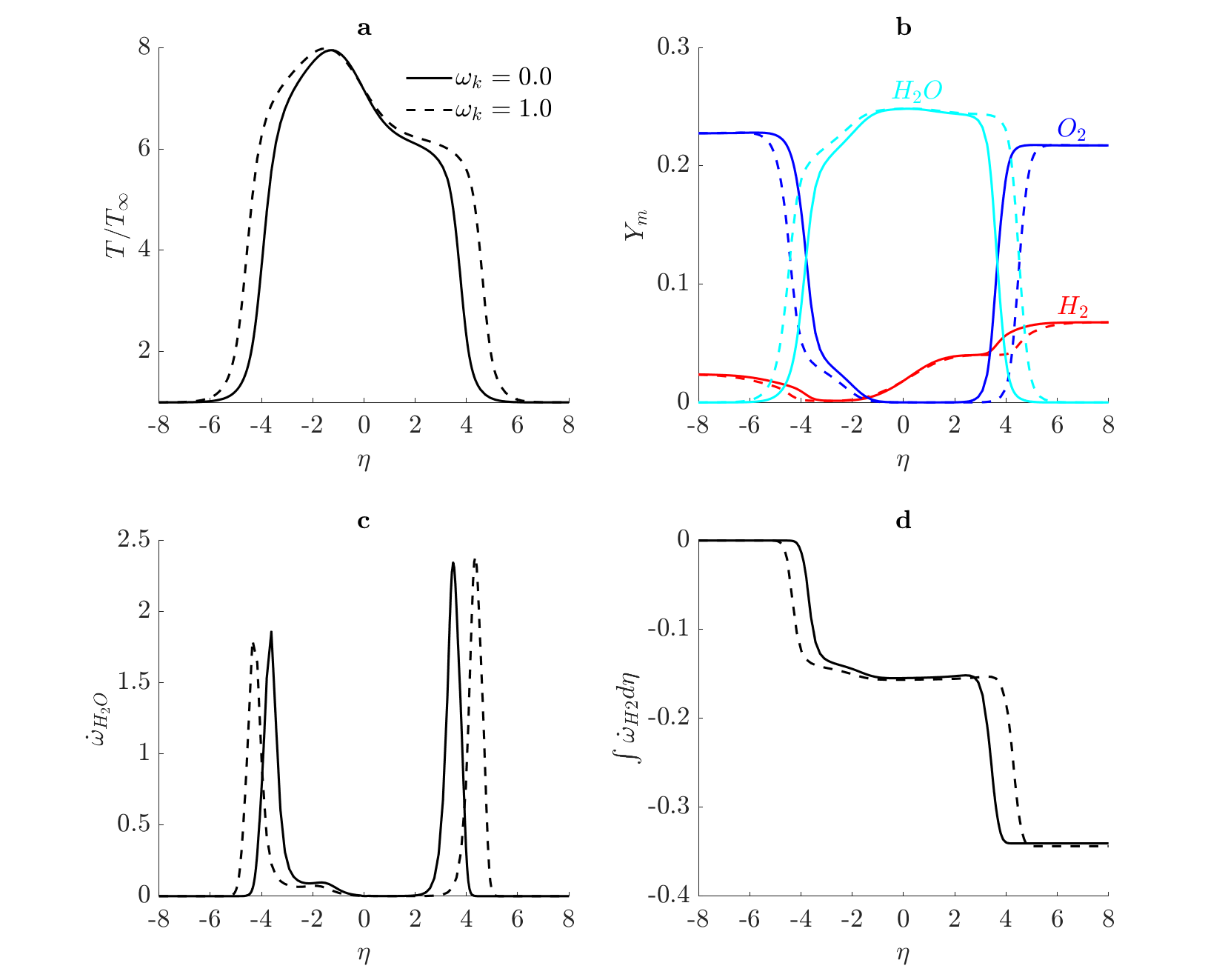}}
\caption{Partially-premixed flame. (a) Temperature; (b) Major species mass fractions; (c) \ce{H2O} reaction rate; (d) Integrated $H_2$ reaction rate (burning rate). $S_1 = 0.5$; $S_2 = 0.5$; $\omega_k = 0.0, 1.0$; $S^* = 30,000$ $1/s$.}
\label{fig16}
\end{figure}
The S-curves of Fig. \ref{fig15} show the vorticity effect is consistent with partially-premixed flames, increasing the extinction limit by 9.02 $\%$. The extinction limit is an order of magnitude lower than the diffusion and premixed flames previously presented due to the high concentration of nitrogen in both streams. The scalar profiles in Fig. \ref{fig16}\textbf{a} and \ref{fig16}\textbf{b} show behavior reminiscent of both diffusion and premixed flames. With vorticity, the flame standoff distance increases as does the peak temperature, particularly at high strain rates. As the strain rate increases, the premixed flames on either side of the stagnation point are pushed into the non-premixed flame and it becomes difficult to distinguish three separate flames. When the strain rate becomes very high, little to no premixed flame exists and the profiles resemble that of the non-premixed flames in Section 3.1. This is also the case across the entire unstable branch where, even at low strain rates, premixed flames do not appear on either side of the non-premixed flame. We believe this is due to insufficient activation energy associated with the low peak temperatures on the unstable branch; i.e., the non-premixed flame does not produce enough energy to sustain the premixed flames.

Figure \ref{fig16}\textbf{c} shows the reaction rate of \ce{H2O} for the partially-premixed flame at a low strain rate. Two premixed flames are quite evident here while the central non-premixed flame is minimal. The centrifugal effect shifts the location of the premixed flames away from the stagnation point according to the modified axial velocity profile. The shape of integrated burning rate profiles in Fig. \ref{fig16}\textbf{d} differs from that of pure diffusion and pure premixed flames as individual `steps' are seen for the two premixed flames, with a slight tilting of the first step due to the central non-premixed flame. Despite the difference in shape, the overall finding that the centrifugal effect decreases burning rate is consistent with pure diffusion and pure premixed flames.

\begin{figure}[!htb]
\centering
\noindent\makebox[\textwidth]{%
\includegraphics[width=1.22\textwidth]{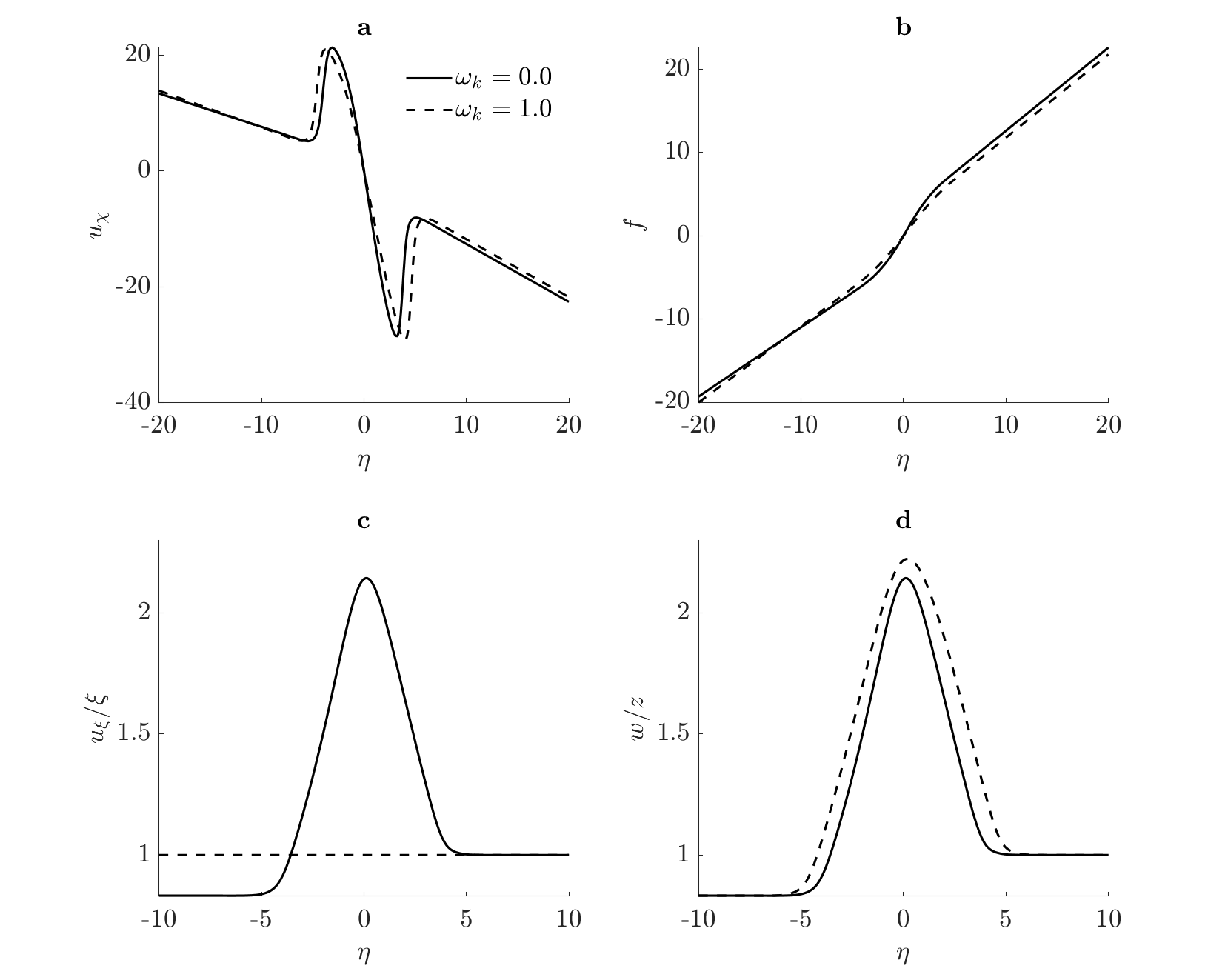}}
\caption{Partially-premixed flame. (a) Axial (inflow) velocity $u_\chi$; (b) Similarity variable $f$; (c) Transverse outflow velocity normal to vorticity vector divided by length $(u_{\xi}/\xi)$; (d) Transverse outflow velocity parallel to vorticity vector divided by length $(w/z)$. $S_1 = 0.5$; $S_2 = 0.5$; $\omega_k = 0.0, 1.0$; $S^* = 30,000$ $1/s$.}
\label{fig17}
\end{figure}
Velocity and mass flux profiles of the premixed flames are presented in Fig. \ref{fig17} and have no qualitative difference from that of the non-premixed flames previously presented. Mass efflux in the $z$ direction is increased by vorticity and there is a corresponding decrease in $\xi$ efflux. The axial strain rate in the flame is decreased and the boundary mass flux is increased on the lean (oxidizer side) and decreased on the rich (fuel side). 

\section{Discussion}
The inclusion of detailed chemical kinetics, mixture-averaged diffusion, and variable thermo-physical properties in the rotational flamelet model has yielded new findings in the effects of vorticity and three-dimensional strain distributions. While corroborating the fundamental result that vorticity increases residence time, making a highly strained flame behave like a moderately-strained, irrotational flame, other differences exist due to reactant composition in concert with centrifugal acceleration. The original demonstration of the rotational flamelet model used oxygen and propane as the reactants, which have, by order of magnitude, comparable molecular weights. Consequently, the location of minimum density coincided with that of peak temperature. Initial testing of the modified model using pure hydrogen and oxygen showed the location of minimum density resides upstream in the hydrogen inflow for non-premixed flames. When vorticity is applied to this case, the tendency of lighter products to exit the domain along the $z$-axis, as identified by \cite{Sirignano1}, is not apparent because the fluid in the flame zone is relatively dense despite the elevated temperature. The non-premixed flame case discussed in Section 3.1, using an equimolar mixture of nitrogen and hydrogen in the fuel stream, is more similar in terms of density to the original propane-oxygen configuration and does show this tendency. Replacing nitrogen with water accomplishes a similar density shift and is qualitatively equivalent. This is also expected in the burning of hydrocarbon fuels.

Key findings:
\begin{itemize}
  \item Applying vorticity to non-premixed, premixed, and partially-premixed counterflow flames increases the flammability limit by reducing the maximum strain rate and thereby increasing residence time. This also reduces the burning rate.
  \item With vorticity, the ratio of the two transverse normal strain rates, $S_{1}f_{1}^{'}$ and $S_{2}f_{2}^{'}$, will vary from the value at the incoming stream. The centrifugal effect can result in a significant turning of the outflow. A three-dimensional flame structure results.
  \item When the transverse strain rate in the direction aligned with the vorticity vector ($S_2$) is larger than the transverse strain rate in the plane normal to the vorticity vector ($S_1$) i.e. ($S_2 > S_1$), flammability limits are extended relative to the opposite inequality of transverse strain rates i.e. ($S_2 < S_1$).
  \item Vorticity effects are maximized when the location of minimum density coincides with the location of peak temperature; i.e., minimum density occurs within the diffusion layer.
  \item The authors believe counterflow geometry and boundary conditions specified by far-field strain rates better approximates resolved scale cell-averaged strain rates while producing velocity and scalar distributions that match nozzle counterflows near the stagnation point and interface.
\end{itemize}

The stable and unstable branches of the extinction curves (S-curves) were calculated, clearly showing the extension of the flammability limit and corroborating the initial findings of \cite{Sirignano1}. Given that vorticity magnitude is largest at the small length scales where flamelet models apply, and that extinction and re-ignition are ubiquitous in turbulent combustion, the precise calculation of extinction behavior is essential. The addition of the vorticity effect to stable- and unstable-branch calculations coupled with the ability to produce three burning regimes more accurately captures this unsteady behavior in turbulent reacting flows.

The improved kinetics, transport formulation, and variable thermo-physical properties also allow qualitative comparison with existing counterflow non-premixed flame codes, such as CHEMKIN OPPDIF and Cantera difflame codes. The key finding of this comparison is that the steady-state velocity solution is critically dependent on the flow geometry. Specifically, when the in-flowing streams are assumed to originate from nozzles with a finite separation distance, a different steady-state solution for velocity is obtained from the rotational flamelet code because nozzles force a zero-velocity gradient at the boundaries. Despite the deviation in velocity near the boundaries, the velocity profiles in the flame zone agree very well for axisymmetric, irrotational flow when the maximum local strain rate is matched. This is the region where scalar gradients exist and thus temperature and mass fraction profiles match well between Cantera and the rotational flamelet model. The use of a potential counterflow to describe the asymptotic incoming streams better represents the physics of turbulent reacting flows because extinction and re-ignition phenomena are dependent on strain rates. In other words, the potential counterflow geometry allows a sub-grid strain rate scaled upward from the resolved-flow strain rate to be imposed as the boundary condition rather than a prescribed mass flux. Adding resolved-flow vorticity compounds the ability of the flamelet model to capture extinction and re-ignition behavior.

Premixed and partially-premixed flame cases were also computed to showcase the ubiquity of the rotational flamelet model. Turbulent reacting flows are not limited to non-premixed flames only and it is thus the belief of the authors that it is important for the flamelet model to handle all flame types. It is also noteworthy that, in comparison to non-premixed flames, the ambient extinction strain rates of the premixed flames are larger by a factor of 2-3 while the ambient extinction strain rates of the partially-premixed flames are smaller by a factor of 4-5. This suggests that flamelet models considering only non-premixed flames may poorly emulate real turbulent flames having all three burning regimes.

This work has focused on the use of detailed transport analysis. Therefore, the scalar dissipation rate, which generally depends on uniform mass diffusivities over all species, has no utility here. A similar analysis, considering three-dimensionality and vorticity for non-premixed flames with a uniform mass diffusivity, is presented by Hellwig et al. \cite{Hellwig2024_1} and produces the same general conclusions about the effects of vorticity and strain rate.

Further work should be aimed at increasing the computational efficiency of the model, implementation of a chemical kinetics model for heavy-hydrocarbon fuels, and production of useful flamelet tables and coupling procedures for use in LES.

\section*{Acknowledgement}
The research was supported by the Air Force Office of Scientific Research through Grant FA9550-22-1-0191 with Dr. Mitat Birkan and more recently Dr. Justin Koo as program manager and by the Office of Naval Research through Grant N00014-21-1-2467 with Dr. Steven Martens as program manager. The Melucci family is thanked for the support of the first author through the William and Ida Melucci Space Exploration \& Technology Fellowship. Discussions with Professor Feng Liu on the numerical methods were valued.

\section*{Disclosure statement}
The authors declare they have no known competing interest that may have influenced the content of this work.

\section*{Biographical Note}
Wes Hellwig is a Ph.D. student in Mechanical and Aerospace Engineering at the Samueli School of Engineering at the University of California, Irvine. His emphasis is in fluid dynamics and propulsion and his research focuses on numerical simulation of turbulent reacting flows. He received his M.S. degree in 2023 from the University of California, Irvine.

Xian Shi is an Assistant Professor at the Samueli School of Engineering at the University of California, Irvine. His research centers around energy conversion and propulsion technology, specifically, high-speed reacting flows, detonation and shock waves, chemical kinetics, and carbon and nanoparticle materials. He received his M.S. degree in 2014 and Ph.D. degree in 2017, both from the University of California, Berkeley.

William Sirignano is a Distinguished Professor at the Samueli School of Engineering at the University of California, Irvine. His vast research portfolio over the past 60 years includes work on combustion theory, computational methods, fluid dynamics, multiphase flows, combustion instability, and propulsion systems. His current research focuses on numerical simulation of practical turbulent reacting flows through Reynolds-Averaged Navier Stokes and Large-Eddy Simulations and high-performance upgrades to gas-turbine engines. He received his M.A. degree in 1962 and Ph.D. degree in 1964, both from Princeton University.

\section*{Data Availability}
Data is available on request from the authors. Please contact the corresponding author. 

\bibliographystyle{tfq}
\bibliography{interactbibliography}

\end{document}